\documentclass[a4paper,11pt]{article}

\usepackage{jcappub} 
\usepackage[T1]{fontenc}
\usepackage[utf8]{inputenc}
\usepackage{amsmath}
\usepackage[capitalize]{cleveref}

\usepackage[dvipsnames]{xcolor}
\usepackage{mathtools}
\usepackage{epstopdf}
\usepackage[section]{placeins}
\usepackage[normalem]{ulem}
\usepackage{xcolor}
\usepackage{cleveref}

\newcommand{\lsim}{\mathrel{\mathop{\kern 0pt \rlap
  {\raise.2ex\hbox{$<$}}}
  \lower.9ex\hbox{\kern-.190em $\sim$}}}
\newcommand{\gsim}{\mathrel{\mathop{\kern 0pt \rlap
  {\raise.2ex\hbox{$>$}}}
  \lower.9ex\hbox{\kern-.190em $\sim$}}}

\newcommand{\lavg}{\ensuremath{\left\langle}}
\newcommand{\ravg}{\ensuremath{\right\rangle}}

\newcommand*{\avg}[1]{\ensuremath{\lavg #1 \ravg}}

\title{Cosmology of Inelastic Self-Interacting Dark Matter: Linear Evolution and Observational Constraints}
\author[a,b]{Xin-Chen Duan,}
\author[a,b]{Yue-Lin Sming Tsai}
\author[c,a]{and Ziwei Wang}

\affiliation[a]{Key Laboratory of Dark Matter and Space Astronomy,\\
Purple Mountain Observatory, Chinese Academy of Sciences, Nanjing 210008, China}

\affiliation[b]{School of Astronomy and Space Science,\\
University of Science and Technology of China, Hefei, Anhui 230026, China}

\affiliation[c]{Department of Strategic and Advanced Interdisciplinary Research, Pengcheng Laboratory, Shenzhen, Guangdong 518066, China}

\emailAdd{xcduan@pmo.ac.cn}
\emailAdd{smingtsai@pmo.ac.cn}
\emailAdd{wangzw@pcl.ac.cn}

\abstract{
We study the linear cosmological evolution of inelastic self-interacting dark matter in a two-component dark sector with a small mass splitting, assuming thermal initial conditions for the two species.  
We derive the coupled background and perturbation equations for inelastic conversion between the two species, 
considering both power-law and low-velocity saturation cross sections. 
Exothermic conversion injects kinetic energy into the light component, generating pressure support that suppresses small-scale structure and produces dark acoustic oscillations in the matter power spectrum. 
The resulting cutoff at scale $k > 1\,h\,\mathrm{Mpc}^{-1}$ depends on the normalization and velocity dependence of the cross section, 
the dark matter mass and the mass splitting. 
Using linear power spectra computed with a modified Boltzmann solver, 
we recast constraints from Lyman-$\alpha$ forest data and high-redshift UV luminosity functions, 
finding non-monotonic but closed exclusion regions driven by the competition between efficient conversion and rapid depletion of the heavy component. 
These results show that the internal thermodynamics of a secluded multi-component dark sector can leave observable imprints on structure formation, providing a complementary probe of secluded dark matter.
}

\begin{document}
\maketitle
\flushbottom

\section{Introduction \label{sec:intro}}

The identity of dark matter (DM) remains one of the most important open questions in modern physics. 
On cosmological scales, the standard collisionless Cold Dark Matter (CDM) paradigm, 
as a cornerstone of the $\Lambda$CDM model, has achieved remarkable success in explaining the cosmic microwave background and the large-scale structure of the Universe~\cite{White:1977jf,Blumenthal:1984bp,Navarro:1995iw}. 
However, as observational precision increases, two distinct lines of evidence from particle physics and astrophysics suggest that 
the dark sector is likely more complex than a single, inert, and collisionless particle.

First, from a particle physics perspective, the null results from direct detection experiments~\cite{XENON:2023cxc,PandaX:2024qfu, DarkSide-50:2022qzh,CDEX:2018lau} and collider searches~\cite{Boveia:2018yeb, Kahlhoefer:2017dnp, Buchmueller:2017qhf} challenge the traditional Weakly Interacting Massive Particle scenario. 
Despite DM direct detection sensitivities reaching the neutrino floor, the lack of an observed signal strongly suggests that DM interactions with the Standard Model (SM) are highly suppressed. This motivates the exploration of hidden or secluded dark sectors~\cite{Pospelov:2007mp, Arkani-Hamed:2008hhe}, where DM is thermally decoupled from the SM and governed by its own internal gauge dynamics.
If the dark sector is secluded and complex, it generically features a spectrum of states rather than a single particle. 
This mass gap fundamentally alters the interaction dynamics. 
However, the cosmological evolution of such sectors imposes stringent constraints. 
Given the tiny DM-SM interaction mandated by DM direct detection, 
the standard thermal paradigm for secluded sectors relies heavily on strong self-interactions to maintain thermal equilibrium within the dark sector itself~\cite{Pospelov:2007mp, Hochberg:2014dra}. 
To avoid suppressing the matter power spectrum through the injection of dark radiation from the decay of heavier states, 
the mass splitting must be kept small~\cite{Cyr-Racine:2015ihg, Poulin:2016nat}. 
In this nearly-degenerate regime, if the dark sector maintains thermal equilibrium through co-annihilation rather than SM couplings, 
the surviving parameter space is tightly squeezed by Big Bang Nucleosynthesis bounds and collider missing-energy searches~\cite{Boehm:2013jpa, Sabti:2019mhn,Tsai:2019eqi,Banerjee:2016hsk,Fan:2022dck}.
As a result, the viable cosmological framework naturally motivates a multi-component dark sector with rich internal thermodynamics but negligible SM interactions.

Second, from an astrophysical perspective, observations on galactic and sub-galactic scales reveal structural anomalies that are difficult to reconcile with purely collisionless dynamics. 
CDM $N$-body simulations predict steep central density profiles in halos, leading to the core-cusp problem in dwarf galaxies~\cite{Flores:1994gz, Moore:1994yx} and the closely related Too-Big-To-Fail problem in massive subhalos~\cite{Boylan-Kolchin:2011qkt} (for a review, see \cite{Bullock:2017xww}).
Furthermore, galaxies with similar maximum circular velocities can exhibit markedly different inner rotation curve shapes, a spread that neither CDM-only simulations nor baryonic feedback can fully reproduce~\cite{Oman:2015xda, Yang:2023jwn, Yang:2020iya}. 
Self-Interacting Dark Matter (SIDM) provides a compelling and fundamental physics alternative that can naturally address both the central density and diversity problems~\cite{Spergel:1999mh, Kaplinghat:2015aga}.

In the SIDM framework, scattering between DM particles allows for heat transfer within the halo, thermalizing the inner regions and naturally producing the observed cored profiles. However, a successful SIDM model must navigate a tight observational needle: solving dwarf galaxy anomalies requires a large self-interaction cross-section, while observations of merging galaxy clusters impose strict upper bounds on the scattering rate~\cite{Tulin:2017ara}. Consequently, the cross-section must be velocity-dependent, which can be naturally achieved by introducing a light mediator~\cite{Feng:2009hw, Loeb:2010gj, Tulin:2013teo, Ghosh:2021wrk}. However, such light-mediator models generically predict sizable DM--nucleus scattering rates, which are constrained by direct detection experiments~\cite{Del_Nobile:2015uua,Kaplinghat:2013yxa}.
Multi-component or inelastic SIDM models offer an elegant solution to this tension and provide additional phenomenological advantages~\cite{Tucker-Smith:2001myb,Finkbeiner:2007kk,Blennow:2016gde}. By introducing a mass splitting between DM states, DM--nucleus scattering becomes an endothermic process, kinematically forbidden at the low velocities characteristic of the Galactic halo, thereby naturally suppressing direct detection signals while preserving large self-interaction cross-sections~\cite{Blennow:2016gde}. 
Moreover, the dissipative nature of inelastic scattering can accelerate gravothermal core collapse in cluster subhaloes, 
potentially alleviating the galaxy–galaxy strong-lensing anomaly~\cite{Meneghetti:2020yif}, 
in which CDM simulations significantly underpredict the observed lensing cross sections of cluster substructures~\cite{Essig:2018pzq, Yang:2021kdf}.

The discussion above motivates a broad class of inelastic and multi-component dark sectors. 
Here we consider a minimal dark-sector configuration: 
two non-relativistic species that are nearly degenerate, long-lived, and interconvert via internal conversion reactions.
This setup captures the redistribution of abundance and kinetic energy within the dark sector while avoiding additional assumptions about sizable dark radiation or Standard-Model interactions. 
Such two-state systems arise naturally in inelastic or excited DM, pseudo-Dirac DM, and dark-sector multiplets~\cite{Tucker-Smith:2001myb,TuckerSmith:2004jv,Finkbeiner:2007kk,Chang:2008gd,Arkani-Hamed:2008hhe,Schutz:2014nka,Blennow:2016gde,Filimonova:2022pkj}. 
We describe these scenarios by a light component $\chi_l$ and a heavy component $\chi_h$ connected through $\chi_h\chi_h\leftrightarrow\chi_l\chi_l$, 
and study how this internal conversion shapes their abundances, temperatures, and perturbations.

Similar internal dark-sector dynamics has been widely studied in the context of nonlinear halo evolution. Dissipative SIDM can accelerate gravothermal evolution through energy loss in inelastic collisions~\cite{Essig:2018pzq,Huo:2019dhalos,Shen:2021fire1,Shen:2022fire2}, while inelastic SIDM with nearly degenerate internal states can modify satellite evaporation, halo profiles, and substructure through excitation and de-excitation processes~\cite{Vogelsberger:2018evap,Chua:2020mw,ONeil:2022endothermic}. More broadly, multi-component models show that several dark species can alter halo density profiles, substructure abundances, and the satellite mass function~\cite{Semenov:2013multicomp,Todoroki:2017substructure,Todoroki:2017profiles,Todoroki:2020clusters,Yang:2025xsp,Yang:2025dgl}. These studies underscore the importance of internal dark-sector dynamics, but they focus on nonlinear halo evolution and typically initialize from a standard CDM power spectrum, thereby neglecting the imprint of dark-sector interactions on linear structure formation.

Our work also connects to linear-regime studies of interacting dark sectors. 
DM interactions with relativistic hidden-sector particles can induce collisional damping and dark acoustic oscillations in the matter power spectrum, as studied in atomic DM models and in the ETHOS framework~\cite{Cyr-Racine:2012tfp,Cyr-Racine:2013fsa,Cyr-Racine:2015ihg,Vogelsberger:2015gpr,Bose:2018juc,Yuan:2026zlr}. Related work has examined the cosmological impact of SIDM and dark radiation on the CMB and matter power spectrum~\cite{Huo:2017vef,Egana-Ugrinovic:2021gnu}, and DM scattering with Standard-Model particles---baryons, photons, electrons, and neutrinos---has been extensively constrained using CMB anisotropies, Lyman-$\alpha$ forest data, and the linear matter power spectrum~\cite{Dvorkin:2013cea,Wilkinson:2013kia,Wilkinson:2014ksa,Gluscevic:2017ywp,Boddy:2018kfv,Xu:2018efh,Boddy:2018wzy,Nguyen:2021cnb,Dhyani:2025uof,Buen-Abad:2021mvc, Zu:2023rmc, Zhang:2024mmg, He:2025npy, Zu:2025lrk, Trojanowski:2025oro,Mosbech:2026xcc}. Recent work on boosted DM~\cite{Kim:2024bdm} has further shown that the annihilation of a heavier dark species into a lighter one can suppress the small-scale power spectrum and soften galactic density profiles in a manner analogous to warm dark matter (WDM), while evading conventional WDM mass constraints.

The present work is complementary to these scenarios. 
Although the linear imprints of various interacting dark sectors 
(including atomic DM, ETHOS-like models, and DM--SM scattering) have been well characterized, 
the corresponding effects of nearly degenerate multi-component DM with internal conversion remain unexplored.
We consider a regime where the dark sector is kinetically decoupled from the SM, 
and $\chi_h\chi_h\leftrightarrow\chi_l\chi_l$ conversions leave an imprint on structure formation 
without violating current self-interaction bounds~\cite{Tulin:2017ara,Eckert:2022qia, Adhikari:2024aff,ODonnell:2025pkw,Jee:2026qjg}.
This setup isolates the imprint of multi-component dark-sector thermodynamics on the homogeneous background and the linear matter power spectrum, 
without invoking sizable dark radiation or SM interactions.

The paper is organized as follows. In Sec.~\ref{sec:cosmo_evol}, 
we summarize the parameterization of the velocity-dependent cross-section for inelastic SIDM. 
We then derive the relevant Boltzmann and Einstein equations governing the thermal history of the universe in Sec.~\ref{sec:homo} 
and linear cosmological perturbations in Sec.~\ref{sec:linearForm}. 
In Sec.~\ref{sec:constraints}, we utilize observations of the Lyman-$\alpha$ forest and the UV luminosity function to impose up-to-date constraints on the model. Finally, we summarize our findings and discuss future directions in Sec.~\ref{sec:summary}.

\section{Evolution equations of inelastic DM}
\label{sec:cosmo_evol}

In this section, we present the theoretical framework for the cosmological evolution of inelastic DM. 
We first introduce the inelastic DM model and the conversion processes between DM particles relevant for cosmology. 
Henceforth, we refer to these conversion processes simply as \textit{``conversion''} unless stated otherwise.
We then describe the impact of conversion  on the homogeneous background evolution. 
Finally, we derive the linear perturbation equations governing the growth of density and velocity fluctuations, 
which form the basis for the cosmological predictions discussed in subsequent sections.

\subsection{Inelastic DM frameworks and their conversion  cross section}
\label{sec:models}
We consider a DM sector composed of a light component $\chi_l$
and a heavy component $\chi_h$, with masses $m_l$ and $m_h$ respectively.
Their mass splitting is $\Delta m = m_h - m_l$, and we require $\Delta m < m_l$. 
Such systems naturally arise within inelastic or excited DM frameworks~\cite{Tucker-Smith:2001myb,TuckerSmith:2004jv,Finkbeiner:2007kk,Chang:2008gd}, as pseudo-Dirac fermions with a small Majorana mass term~\cite{DeSimone:2010,Konar:2020wvl,Konar:2020vuu}, or as multiplets within a non-Abelian dark sector~\cite{Arkani-Hamed:2008hhe,Baumgart:2009,Katz:2009qq,Chen:2009ab,Zhang:2009dd}.
Our analysis focuses on the inelastic conversion process\footnote{We neglect elastic scattering $\chi_h\chi_l \leftrightarrow \chi_h\chi_l$ and retain only the conversion reaction $\chi_h \chi_h \leftrightarrow \chi_l \chi_l$, as it is the source of non-trivial cosmological effects. 
Such elastic scattering can redistribute momentum between the two components and smooth small-scale features induced by the conversion reaction (e.g., softening sharp cutoffs or oscillatory structure), without qualitatively altering our conclusions.} $\chi_h \chi_h \Leftrightarrow \chi_l \chi_l$, 
in which the mass splitting is converted into (or drawn from) kinetic energy.
We assume that self-interactions within each component sufficiently maintains approximate kinetic equilibrium\footnote{The elastic self-interactions required to maintain kinetic equilibrium can
remain consistent with cluster constraints, for example through a
velocity-dependent cross section that is suppressed at cluster velocities. 
Moreover, in the non-relativistic regime considered here, they do not generate
additional explicit collision terms in the number density or velocity
perturbation equations, but instead suppress higher moments~\cite{Cyr-Racine:2015ihg}.}.

We parameterize the DM conversion  cross section $\sigma$ for the $\chi_h \chi_h\Leftrightarrow \chi_l \chi_l$ in two forms.  In the limit $\Delta m / m_l \ll 1$, the  exothermic process  $\chi_h\chi_h \to \chi_l\chi_l$ and endothermic process $\chi_l\chi_l \to \chi_h\chi_h$ cross sections are approximately equal.

\begin{itemize}
    \item \textbf{Power-Law Scaling:} The conversion  cross section is parameterized by the velocity-dependent relation 
        \begin{equation}
            \sigma_{\chi_h\chi_h \Leftrightarrow \chi_l\chi_l} \equiv \sigma_{\rm PL} \left(\frac{v_{\rm rel}}{c}\right)^{n-1}
            \label{eq:sigma_powerlaw}
        \end{equation}
where $\sigma_{\rm PL}$ denotes the effective interaction strength and $n$ follows the partial-wave expansion.
The case $n=0$ corresponds to $s$-wave contact interactions from short-range couplings.
For $n=-1$, the cross section  corresponds to the standard Sommerfeld enhancement ($\sigma v \propto v^{-1}$), a type of long-range forces mediated by light bosons~\cite{Arkani-Hamed:2008hhe}. 
Finally, the steep $n=-2$ scaling describes resonant Sommerfeld enhancement near a zero-energy pole~\cite{Iengo:2009ni},
serving as a phenomenological benchmark for maximal velocity enhancement.

\item \textbf{Low-velocity saturation:}
We consider a velocity-dependent cross section whose corresponding $\sigma v_{\rm rel}$ approaches a constant plateau as $v_{\rm rel} \to 0$, 
capturing the saturation behavior expected in Sommerfeld-enhanced interactions. 
Unlike a pure power-law parameterization, which yields a divergent interaction rate at low velocities, 
the scattering amplitude is bounded by unitarity~\cite{Griest:1989wd} and by the finite range of the force~\cite{Cassel:2009wt}, 
leading to saturation at low velocities.
To incorporate this behavior while maintaining analytical simplicity, 
we adopt a regularized form that interpolates between power-law scaling at high velocities and a constant
$\sigma v_{\rm rel}$ at low velocities,
\begin{equation}
    \sigma_{\chi_h\chi_h \Leftrightarrow \chi_l\chi_l} \equiv 
    \frac{\sigma_{\rm sat} (c/v_{\rm rel})
    }{1 + (v_{\rm rel}/v_0)^4}
\label{eq:sigma_saturated}
\end{equation}
where $v_0$ denotes the characteristic velocity and $\sigma_{\rm sat}$ sets the low-velocity limit.
This functional form provides a standard analytic fit to the cross section induced by a Yukawa potential 
and offers a simple two-parameter characterization of the saturation behavior, 
allowing us to systematically explore its impact on structure formation.
\end{itemize}

\begin{table}[t]
    \centering
    \label{tab:BP}
    \begin{tabular}{cccc}
        \hline\hline
        Benchmark & $\sigma_{\chi_h\chi_h \Leftrightarrow \chi_l\chi_l}$ & velocity dependence & Normalization \\
        \hline
        BP1 & 
        Eq.~\eqref{eq:sigma_powerlaw} & $n=0$  & $\sigma_{\rm PL} = 10^{-31}$ (cm$^2$)\\
        BP2 &  Eq.~\eqref{eq:sigma_powerlaw}                                         & $n=-1$ & $\sigma_{\rm PL} = 10^{-36}$ (cm$^2$)\\
        BP3 &  Eq.~\eqref{eq:sigma_powerlaw}                                         & $n=-2$ & $\sigma_{\rm PL} = 10^{-43}$ (cm$^2$)\\
        \hline
        BP4 &  
        Eq.~\eqref{eq:sigma_saturated} & Low-velocity saturation   & $\sigma_{\rm sat} = 10^{-28}$ (cm$^2$)\\
        \hline\hline
    \end{tabular}
    \caption{Benchmark points adopted in this work.
The normalizations for BP2 to BP4 are set at the center of the exclusion regions derived in Sec.~\ref{sec:paramter_space}, while 
BP1 uses a normalization intended only for comparison.
All benchmarks adopt $m_\ell = 100\,\mathrm{MeV}$ and $\Delta m/m_\ell = 10^{-2}$.}    
\end{table}

Given the distinct velocity structures, a single normalization cannot produce comparable effects for all scenarios. 
In Table~\ref{tab:BP}, we define a set of representative cross-section benchmarks designed to illustrate the phenomenological impact of each velocity structure.
We set normalizations at the center of the exclusion regions derived in Sec.~\ref{sec:paramter_space}, for velocity-enhanced cross-section ($n = -1, -2$, and $\sigma_{\rm sat}$). 
In comparison, we also include the $s$-wave cross-section (BP1).~\footnote{We adopt an arbitrary normalization of $\sigma_{\rm PL} = 10^{-31}$~cm$^2$, 
since the $s$-wave conversion between dark sectors alters cosmological linear perturbations negligibly.} 
The mass parameters are set to $m_\ell = 100\,\mathrm{MeV}$ and $\Delta m/m_\ell = 10^{-2}$ for all benchmarks.

\subsection{Background evolution Equations}

The cosmological evolution of the DM is governed by interactions between the heavy and light species.
The redshift evolution of the heavy component fraction $r_h$ satisfies
\begin{equation}
    \frac{d r_h}{d z}
    = \frac{1}{H(z)} \left[ r_h \mathcal{R}_d - (1 - r_h) \mathcal{R}_u \right],
    \label{eq:r_h}
\end{equation}
where $H(z)$ is the Hubble parameter and
\begin{equation}
    r_h(z) \equiv \frac{n_h(z)}{n_l(z) + n_h(z)},
\end{equation}
with $n_l$ and $n_h$ denoting the number densities of the light and heavy species, respectively.
Here $\mathcal{R}_d$ and $\mathcal{R}_u$ denote the conversion 
rates for the exothermic process $\chi_h \chi_h \to \chi_l \chi_l$ and the endothermic process $\chi_l \chi_l \to \chi_h \chi_h$ in the collision term,
\begin{equation}
    \mathcal{R}_d
    \equiv \frac{1}{1+z} \frac{\bar{\rho}_{\chi_h}}{m_{\chi_h}}
    \langle \sigma v \rangle_{\chi_h \chi_h \to \chi_l \chi_l},
    \qquad
    \mathcal{R}_u
    \equiv \frac{1}{1+z} \frac{\bar{\rho}_{\chi_l}}{m_{\chi_l}}
    \langle \sigma v \rangle_{\chi_l \chi_l \to \chi_h \chi_h},
\end{equation}
where $\bar{\rho}_{\chi_i}$ is the background energy density and $\langle \sigma v \rangle$ denotes the velocity-weighted thermally averaged cross section.
The explicit expressions for all conversion rates are given in Appendix~\ref{sec:num_density}.
The conversion between heavy and light species is asymmetric. The endothermic process $\chi_l \chi_l \to \chi_h \chi_h$ is kinematically suppressed by the threshold $v_{\rm rel} > c\sqrt{8\,\Delta m/m_l}$, compared to the exothermic process $\chi_h \chi_h \to \chi_l \chi_l$.

The temperatures of the light and heavy components, $T_l$ and $T_h$, evolve as
\begin{equation}
    \frac{d T_l}{d z}
    = 2 \frac{T_l}{1+z}
    - \frac{2}{3} \frac{\mathcal{R}_d}{H(z)}
    \left[ \Delta m + \frac{3}{2}(T_h - T_l) \right],
    \label{eq:T_l}
\end{equation}
\begin{equation}
    \frac{d T_h}{d z}
    = 2 \frac{T_h}{1+z}
    - \frac{2}{3} \frac{\mathcal{R}_u}{H(z)}
    \left[ -\Delta m + \frac{3}{2}(T_l - T_h) \right].
    \label{eq:T_h}
\end{equation}
In both equations, the first term on the right-hand side describes adiabatic cooling due to Hubble expansion, while the second term accounts for heat exchange from particle conversions. The exothermic process deposits kinetic energy $\Delta m$ per conversion into the light species, whereas the endothermic process extracts it from the light population.

\cref{eq:r_h,eq:T_l,eq:T_h} together determine the homogeneous thermochemical evolution of the dark sector, including departures of the heavy and light species from chemical and thermal equilibrium.
The derivations of the number density and temperature evolution equations are presented in Appendices~\ref{sec:num_density} and~\ref{sec:temp_evolution}, respectively.
The resulting redshift-dependent quantities $r_h(z)$, $T_l(z)$, and $T_h(z)$ provide the necessary background inputs for the linear perturbation equations discussed in the next section.

\subsection{Perturbation Equations}

To describe the growth of structure, we work in the Newtonian gauge and consider linear perturbations.
The density perturbations of the two species, $\delta_l$ and $\delta_h$, evolve as
\begin{equation}
    \delta_l'
    = -\theta_l + 3\Phi'
    + \xi \mathcal{R}_{d} \bigl(2\delta_h - \delta_l + \Psi \bigr)
    - \mathcal{R}_{u} \bigl( \delta_l + \Psi \bigr),
    \label{eq:delta_l}
\end{equation}
\begin{equation}
    \delta_h'
    = -\theta_h + 3\Phi'
    + \frac{1}{\xi}\mathcal{R}_{u} \bigl(2\delta_l - \delta_h + \Psi \bigr)
    - \mathcal{R}_{d} \bigl( \delta_h + \Psi \bigr),
    \label{eq:delta_h}
\end{equation}
where a prime ($\prime$) denotes a derivative with respect to conformal time $\tau$, $\Psi$ and $\Phi$ are the Bardeen potentials in the Newtonian gauge, 
$\theta$s are the velocity divergences, and
$\xi \equiv \bar\rho_h/\bar\rho_l$.
The corresponding momentum conservation equations are
\begin{equation}
    \theta_l'
    = -\mathcal{H}\theta_l + k^2 \Psi
    + k^2 \hat{c}_{s,l}^2 \delta_l
    + \xi \mathcal{R}_{d} \bigl(\theta_h - \theta_l \bigr),
    \label{eq:theta_l}
\end{equation}
\begin{equation}
    \theta_h'
    = -\mathcal{H}\theta_h + k^2 \Psi
    + \frac{1}{\xi}\mathcal{R}_{u} \bigl(\theta_l - \theta_h \bigr),
    \label{eq:theta_h}
\end{equation}
where $\mathcal{H} \equiv aH$ is the conformal Hubble parameter and $\hat{c}_{s,l}^2$ is the rest-frame sound speed of the light component. The derivation of the perturbation equations is presented in Appendix~\ref{app:perturb_derivation}.
Under the assumption of local thermal equilibrium, we identify $\hat{c}_{s,l}^2$ with the adiabatic sound speed $c_{a,l}^2 \equiv \dot{\bar{P}}_l / \dot{\bar{\rho}}_l$. 
The detailed definition of sound speed and justification of this approximation are given in Appendix~\ref{app:sound_speed}.
Note that the pressure term $k^2 \hat{c}_{s,h}^2 \delta_h$ has been omitted from Eq.~\eqref{eq:theta_h}, as $T_h$ is governed solely by the endothermic process $\chi_l\chi_l \to \chi_h\chi_h$, whose cross section is kinematically suppressed at early times. 
Hence, $T_h$ evolves through adiabatic cooling, yielding $\hat{c}_{s,h}^2 \equiv T_h/m_h \ll \hat{c}_{s,l}^2$, and the corresponding pressure support for the heavy component is negligible.

In the density equations \eqref{eq:delta_l} and \eqref{eq:delta_h}, 
the terms proportional to $\mathcal{R}_d$ and $\mathcal{R}_u$ arise from perturbations in conversion rates and indicate how local density fluctuations affect net particle transfer between species.
In the momentum equations \eqref{eq:theta_l} and \eqref{eq:theta_h}, 
the collision terms act as a drag force that tends to equilibrate the bulk velocities of the two components. The finite temperature of the light species further provides a pressure support through the $\hat{c}_{s,l}^2$ term that resists gravitational collapse. For sufficiently large reaction rates, the combined effect of drag and pressure support suppresses the growth of perturbations, leading to observable deviations in the matter power spectrum relative to the standard $\Lambda$CDM scenario.  
The resulting signatures will be quantified in Sec.~\ref{sec:linearForm}.

To solve the coupled system of background and perturbation equations, 
we implement the two-component DM model into the Boltzmann solver \texttt{CLASS}~\footnote{
We implement our modifications based on \texttt{CLASS}~\cite{Blas:2011rf} (version 3.2.5). The modified code used in this work is publicly available at \url{https://github.com/xcduan-1/2cSIDM-CLASS}.}, incorporating the inelastic conversion dynamics at both the background and perturbation levels.
Unless otherwise stated, the standard cosmological parameters are 
fixed to the \textit{Planck} 2018 best-fit values~\cite{Planck:2018vyg}: 
$H_0 = 67.36\,\mathrm{km/s/Mpc}$, 
$\Omega_b h^2 = 0.02237$, 
$\Omega_c h^2 = 0.1200$, 
$\ln(10^{10}A_s) = 3.044$, 
$n_s = 0.9649$, and 
$\tau_{\rm reio} = 0.0544$, 
where $\Omega_c$ denotes the total DM density. 
In this work, we assume that the DM consists entirely of $\chi_l$ and $\chi_h$, with $\Omega_{\chi_l} + \Omega_{\chi_h} = \Omega_c$.
The dark sector is then characterized by the free parameters $\{m_l$, $\Delta m / m_l\}$, and the cross-section parameters: 
$\{\sigma_{\rm PL},\, n\}$ for the power-law scaling  or $\{\sigma_{\rm sat},\, v_0\}$ for the low-velocity saturation.

\section{Homogeneous Evolution}
\label{sec:homo}
We begin the evolution at the epoch of kinetic decoupling of the dark sector from the SM plasma.
Prior to this epoch, frequent momentum-exchange interactions maintain thermal contact between the dark and visible sectors, so that the dark-sector temperature tracks the photon temperature $T_\gamma$.
At kinetic decoupling, the momentum-transfer rate drops below the Hubble rate, and the dark sector subsequently evolves as a thermally isolated system governed by Eqs.~\eqref{eq:r_h}--\eqref{eq:T_h}.
At this point, both components share a common temperature $T_l = T_h \equiv T_{\rm init}$, 
due to their prior thermal contact with the SM plasma.

We parametrize the initial temperature by $\eta_{\rm init} \equiv m_l / T_{\rm init}$ and adopt the fiducial value $\eta_{\rm init} = 100$.
This ensures that both species are deeply non-relativistic at the onset of evolution, with typical velocities $v \sim \sqrt{T/m} \sim 0.1 c$, 
so that the non-relativistic fluid treatment underlying our Boltzmann equations is well justified.
We note that the precise value of $\eta_{\rm init}$ depends on the details of the dark--SM coupling responsible for kinetic
decoupling~\cite{Bringmann:2006mu}, and can vary over several orders of magnitude among different models. 
We confirm that our results are insensitive to the exact value of $\eta_{\rm init}$, as long as it is large enough to satisfy the non-relativistic condition.

The initial fraction of the heavy component is then given by the Boltzmann ratio,
\begin{equation}
r_{h}(z=z_{\rm init}) = \frac{n_h(z_{\rm init})}{n_h(z_{\rm init}) +n_l(z_{\rm init})}=\left[1 + \exp\!\left(\eta_{\rm init} \frac{\Delta m}{m_l}\right)\right]^{-1}, 
\label{eq:fh_init}
\end{equation}
where we have assumed equal internal degrees of freedom for both components.
The initial redshift $z_{\rm init}$ is determined by matching the initial dark-sector temperature to the photon temperature,
$T_{\rm init} = T_\gamma(z_{\rm init})$
\footnote{For a given light-particle mass $m_l$, the initial temperature is $T_{\rm init} = m_l / \eta_{\rm init}$, and the corresponding redshift follows from entropy conservation,
$z_{\rm init} = \left(g_{*s}(T_{\rm init})/g_{*s,0}\right)^{1/3}
(T_{\rm init}/T_{\gamma,0}) - 1$,
where $g_{*s}(T)$ is the effective number of entropic degrees of
freedom~\cite{Gondolo:1990dk}.}.

\subsection{Reaction Rate and Decoupling}
\label{sec:reaction_rate}
\begin{figure}
    \centering
    \includegraphics[width=\linewidth]{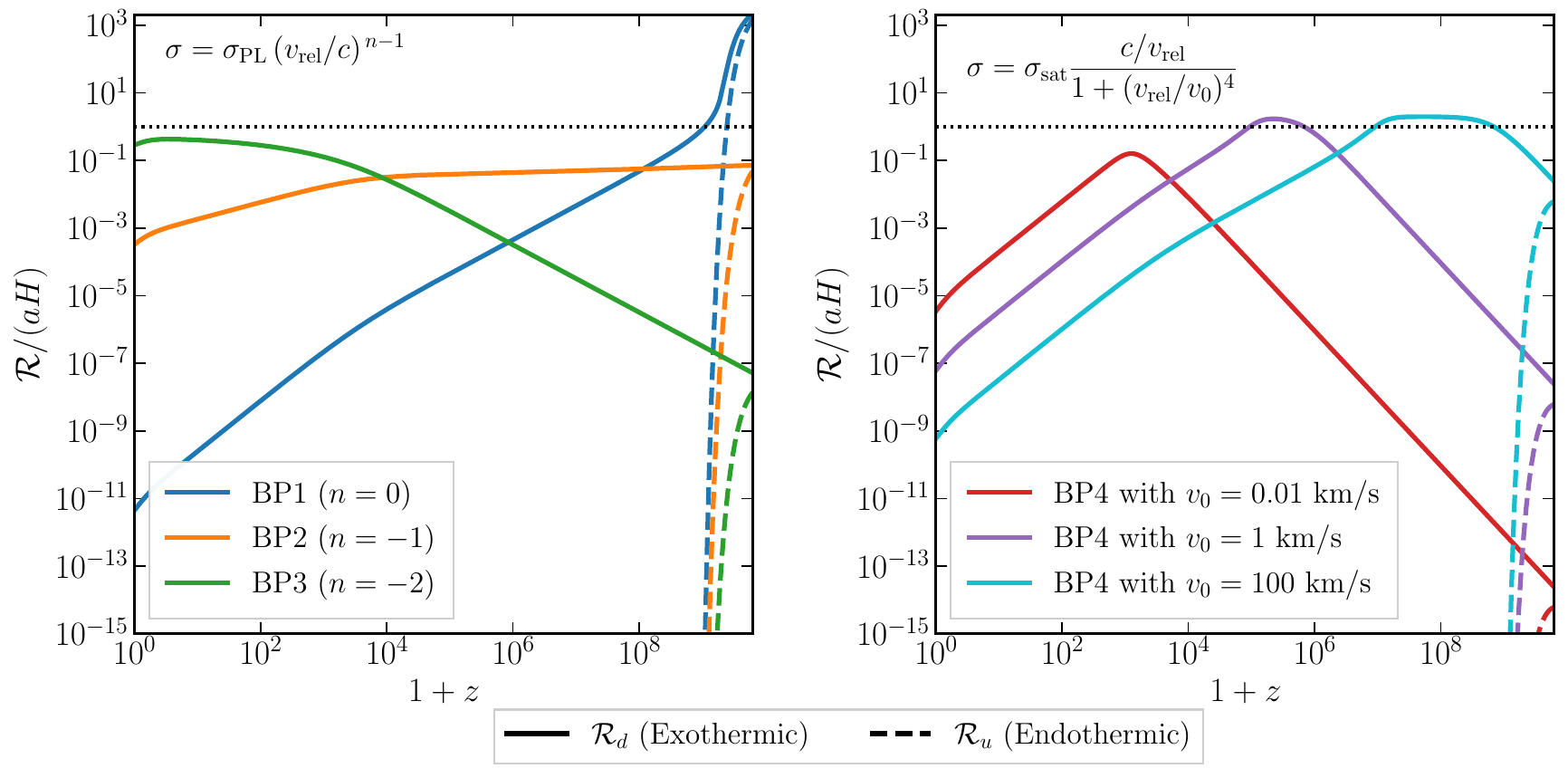}
    \caption{Ratio of the inelastic reaction rate to the Hubble rate, 
    $\mathcal{R}/(aH)$, as a function of $1+z$ for the two 
    cross-section parametrizations considered in this work. 
    Solid lines denote the exothermic (down-scattering) rate 
    $\mathcal{R}_d$, while dashed lines denote the endothermic 
    (up-scattering) rate $\mathcal{R}_u$. 
    The horizontal dotted line marks $\mathcal{R}/(aH) = 1$, 
    above which inelastic conversions proceed faster than the Hubble expansion.
    \textit{Left panel}: Power-law parametrization   
    for the benchmark points 
    BP1 ($n=0$, blue), BP2 ($n=-1$, orange), and BP3 ($n=-2$, green) 
    as listed in table~\ref{tab:BP}. 
    \textit{Right panel}: Saturated parametrization,  
    with $\sigma_{\rm sat} = 10^{-28}\,\mathrm{cm}^2$ (BP4) 
    and three representative saturation velocities for 
    $v_0 = 0.01$ (red), $1$ (purple), and $100\;\mathrm{km/s}$ (cyan). 
    }
    \label{fig:reaction_rate_vs_diffmodel}
\end{figure}

As established by Eq.~\eqref{eq:r_h}-\eqref{eq:T_h}, the evolution of $r_h$ is governed by the competition between conversion reaction rates and cosmic expansion, while the temperature $T_l$ of the light component traces the interplay between adiabatic cooling and kinetic energy injection from exothermic down-scattering. 
For $s$-wave conversion ($n=0$), the light and heavy species remain in thermal equilibrium at early times.
In contrast, for $n < 0$ or the low-velocity saturation case, the conversion rate at early times is suppressed by the large particle velocities.  
Interactions between the light and heavy species are therefore negligible, and both species undergo similar adiabatic cooling from the same initial temperature.
As the universe expands and DM particles slow down, 
the endothermic process becomes exponentially suppressed once the typical kinetic energy drops close to $\Delta m$, and the system is subsequently driven entirely by the exothermic process.

Fig.~\ref{fig:reaction_rate_vs_diffmodel} illustrates the cosmological evolution of the reaction rates relative to Hubble expansion for both the power-law and low-velocity saturation scenarios. 
As anticipated, $\mathcal{R}_u$ (dashed lines) drops sharply at late times, leaving $\mathcal{R}_d$ (solid lines) to dominate the subsequent evolution.

In the case of power-law scaling (left panel), the reaction rate exhibits a dependence on the velocity index $n$. 
Using the benchmark points defined in Table~\ref{tab:BP}, we find that negative indices ($n < 0$) significantly enhance the interaction at low velocities, 
thereby sustaining efficient component conversion down to lower redshifts. 
For $n = -2$, this low-velocity enhancement partially compensates for cosmic dilution, 
producing an extended plateau in the reaction rate that can persist into the structure-formation era. 
By contrast, the $s$-wave case is dominated by early-time dynamics, with the rate rapidly declining at low redshifts.

For the low-velocity saturation case (right panel), the behavior is controlled by $v_0$. 
At early times, when particle velocities are much larger than $v_0$ (namely, $\sigma v \propto v^{-4}$), 
$\mathcal{R}/(aH)$ grows rapidly as particles slow down.
Once $v_{\rm rel} \lesssim v_0$, $\sigma v$ saturates at $\sigma_{\rm sat}$ and the low-velocity enhancement shuts off. 
The ratio $\mathcal{R}/(aH)$ then reaches a maximum and declines as $n_h$ decreases, due to both cosmic dilution and depletion of $r_h$ from exothermic conversion. 
A larger $v_0$ shifts the onset of saturation to higher redshift, and thus moves the peak to earlier times. 
The width of the peak depends on the background cosmology, as transitions during radiation domination (large $v_0$) produce broader features 
than those during matter domination (small $v_0$), owing to the different $H(z)$ scaling in these eras.

\subsection{Chemical and Thermal Evolution}
\label{sec:evolution}
\begin{figure}
    \centering
    \includegraphics[width=\linewidth]{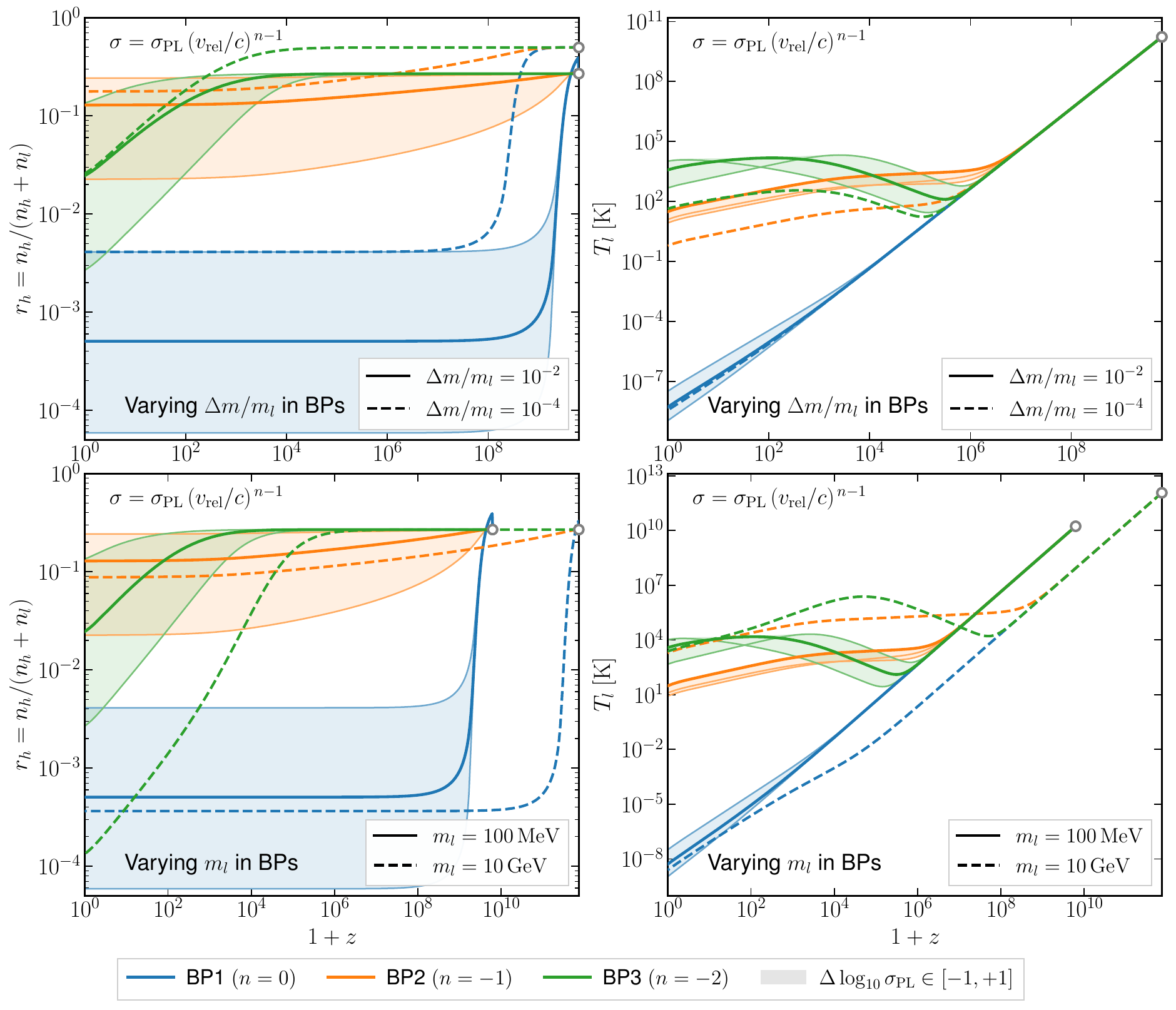}
    \caption{Homogeneous evolution of the heavy-component fraction 
    $r_h = n_h/(n_h+n_l)$ (left column) and the light-component 
    temperature $T_l$ (right column) as a function of $1+z$ 
    for the power-law cross-section parametrization.  
    Blue, orange, and green curves correspond to BP1, BP2, and BP3 given in Table~\ref{tab:BP}. 
    The gray cycles mark the initial redshift $z_{\rm init}$.
    The top row varies the mass splitting 
    $\Delta m/m_l$ at fixed $m_l = 100\,\mathrm{MeV}$, 
    while the bottom row varies $m_l$ at fixed 
    $\Delta m/m_l = 10^{-2}$.  
    Solid lines denote the benchmark values and dashed lines the 
    varied values.  
    Shaded bands indicate a factor-of-ten variation of the 
    cross section around each benchmark, 
    $\Delta \log_{10} \left[\sigma_{\rm PL}\right] \in [-1,\,+1]$.
    }
    \label{fig:powerlaw_fh_Tl}
\end{figure}

We now examine how inelastic conversion shapes the homogeneous evolution of the dark sector, focusing on two observables: the heavy-component fraction $r_h$ and the light-component temperature $T_l$.
\footnote{We do not track $T_h$ separately, as the strong early suppression of $\mathcal{R}_u$ renders the 
conversion term in Eq.~\eqref{eq:T_h} negligible, so that $T_h$ follows pure adiabatic cooling throughout.}
Although both quantities are driven by the same underlying conversion process, their transition epochs generally do not coincide, as they are governed by different criteria.
Once $\mathcal{R}_u$ becomes negligible, Eq.~\eqref{eq:r_h} reduces to $d\ln r_h/dz \approx \mathcal{R}_d/H$, 
while efficient conversion from heavy species to light requires $\mathcal{R}_d \gtrsim aH$.
On the other hand, a departure from adiabatic cooling occurs only when the heating term $(\mathcal{R}_d/H)[\Delta m + \frac{3}{2}(T_h-T_l)]$ in Eq.~\eqref{eq:T_l} becomes comparable to the adiabatic cooling term $2T_l/(1+z)$.
As a result, the depletion of the heavy component and the thermal departure of the light component generally begin at different redshifts.
The velocity dependence of the cross section determines which process occurs first, thereby mapping the homogeneous evolution onto cross-section parameter dependencies.

\begin{figure}
    \centering
    \includegraphics[width=\linewidth]{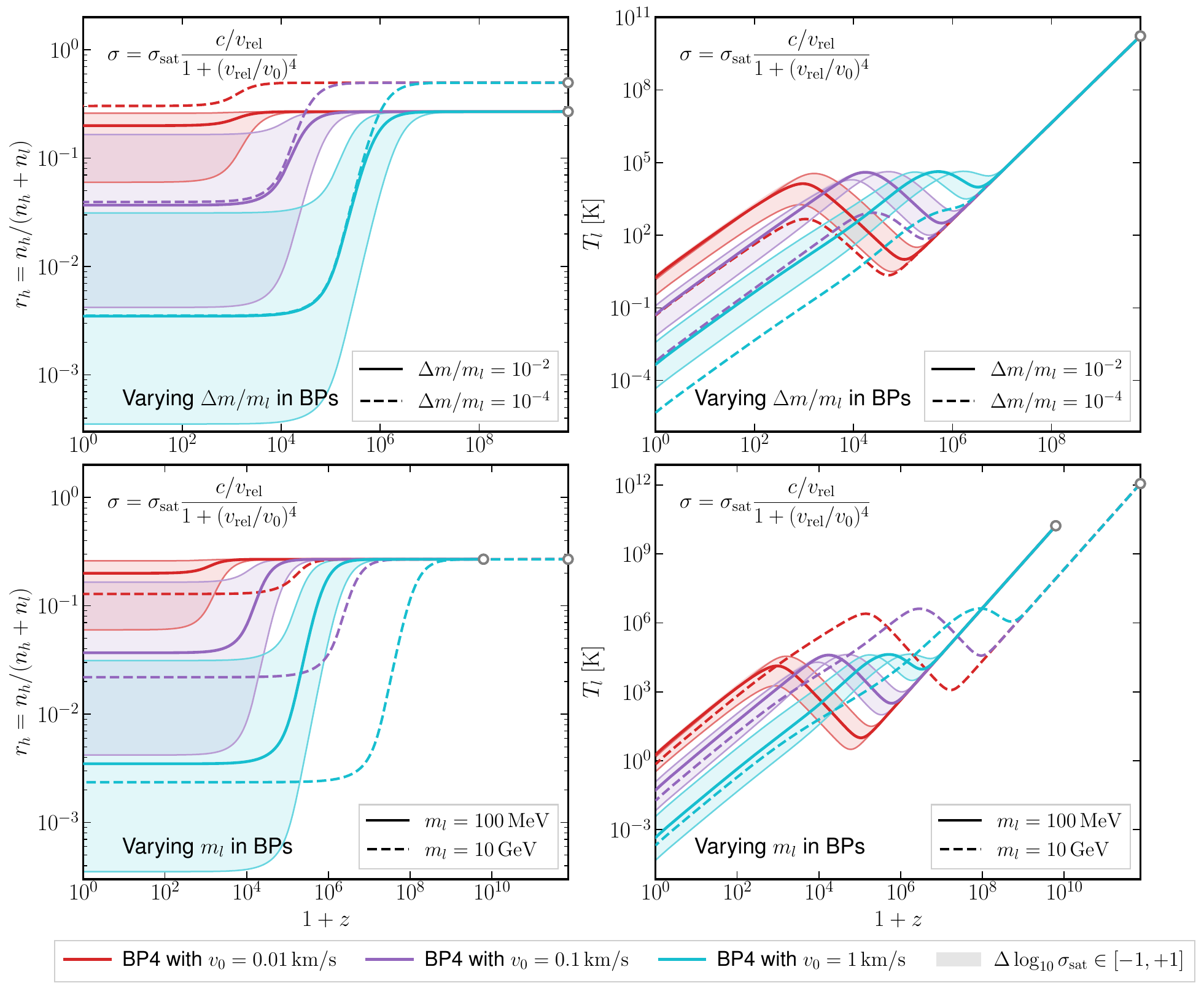}
    \caption{Analogous to Fig.~\ref{fig:powerlaw_fh_Tl} but for the low-velocity saturation cross-section. 
Red, purple, and cyan curves correspond to saturation velocities $v_0 = 0.01$, $0.1$, and $1~\mathrm{km/s}$, respectively. 
We fix $\sigma_{\rm sat} = 10^{-28}~\mathrm{cm}^2$ (BP4). 
Shaded bands represent the cross-section normalization varying by a factor of ten, namely $\Delta \log_{10}\left[\sigma_{\rm sat}\right] \in [-1, +1]$.}
    \label{fig:saturated_fh_Tl}
\end{figure}

Furthermore, the dependence of the homogeneous evolution on these parameters is not purely dynamical. 
The mass splitting $\Delta m/m_l$ not only controls the conversion dynamics but also sets the initial heavy-component fraction through the Boltzmann factor in Eq.~\eqref{eq:fh_init}, 
while $m_l$ determines the initial temperature $T_{\rm init} = m_l/\eta_{\rm init}$ and $z_{\rm init}$. 
Hence, the result of homogeneous evolution depends on the initial conditions and subsequent conversion dynamics, 
as shown numerically in Fig.~\ref{fig:powerlaw_fh_Tl} and Fig.~\ref{fig:saturated_fh_Tl}.

Fig.~\ref{fig:powerlaw_fh_Tl} illustrates these general features for the power-law parametrization. 
The left and right panels show the evolution of $r_h$ and $T_l$, respectively. 
The upper and lower panels correspond to different values of $\Delta m/m_l$ and $m_l$, 
while the shaded bands indicate the effect of the cross-section normalization. 
The initial redshift $z_{\rm init}$ is marked by the gray circles. 
We summarize the key points of Fig.~\ref{fig:powerlaw_fh_Tl} below:

\begin{itemize}
    \item \textbf{Effect of velocity-dependence:} The velocity dependence of the cross section determines the 
ordering of chemical depletion and thermal departure, 
thereby controls the final value of $T_l$.

For $n=0$, depletion occurs before heating. 
The heavy component is mostly exhausted before $T_l$ deviates from adiabatic cooling, 
thus heating is negligible. 
For $n=-2$, heating occurs before depletion. 
In this case $T_l$ deviates from adiabatic cooling while a considerable heavy fraction remains, 
leading to a higher final $T_l$.
The $n=-1$ case lies between these two extremes, with depletion and heating proceeding more concurrently.

    \item \textbf{Role of $\Delta m/m_l$:} The mass splitting plays a dual role. 
A larger $\Delta m/m_l$ releases more energy per conversion, but simultaneously suppresses the initial $r_h$ through the Boltzmann factor in Eq.~\eqref{eq:fh_init}, reducing the total energy budget available for heating.

Comparing $\Delta m/m_l = 10^{-2}$ (solid) and $10^{-4}$ (dashed) in the upper panels, 
the larger splitting produces a higher final $T_l$ for $n=-2$ and $n=-1$. 
For $n=0$ the effect is weaker, as the heavy component is already depleted before heating becomes effective.
For sufficiently large $\Delta m/m_l$, the exponential suppression of the initial $r_h$ eventually dominates, e.g., $r_{h}(z_{\rm init}) \sim 10^{-5}$ for $\Delta m/m_l = 0.1$, leading to 
a non-monotonic dependence of the final $T_l$ on $\Delta m/m_l$.

    \item \textbf{Role of $m_l$:}  
    A larger $m_l$ increases $z_{\rm init}$ and $T_{\rm init}$
    \footnote{
Although both $m_l$ and $\eta_{\rm init}$ enter $T_{\rm init}=m_l/\eta_{\rm init}$,
$\eta_{\rm init}$ is only an initialization parameter once it is large enough
for the system to start in the non-relativistic regime, whereas $m_l$ is a
physical mass scale that affects the conversion history.}, 
    leading to earlier chemical depletion of the heavy component and 
    earlier deviation of the light component from adiabatic cooling.  
    As a result, $r_h$ at a given redshift is smaller than in cases with lower $m_l$.

The dependence of the final $T_l$ on $m_l$ is less straightforward. Comparing $m_l = 100\,\mathrm{MeV}$ (solid) and $10\,\mathrm{GeV}$ (dashed) in the lower panels,
the larger $m_l$ produces a lower final $r_h$ among all scenarios.
For $n=0$, the temperature uplift is small, so varying $m_l$ has 
weak effect on the final $T_l$.
For $n=-2$, a larger $m_l$ extends the effective heating period but simultaneously enhances the depletion of the heavy component. 
These two effects largely compensate, leaving the final $T_l$ only weakly dependent on $m_l$. 
However, this compensation is incomplete for $n=-1$, and the final $T_l$ retains a more pronounced dependence on $m_l$.

    \item \textbf{Impact of $\sigma_{\text{PL}}$:} The cross-section normalization $\sigma_{\rm PL}$ directly controls the overall conversion strength. 

A large $\sigma_{\rm PL}$ accelerates depletion, resulting in a smaller final $r_h$, 
but affects $T_l$ non-monotonically. 
A large $\sigma_{\rm PL}$ can initiate heating earlier but depletes the heavy component faster.
This competition is clearly seen from $n = -2$, 
where the temperature curves for different $\sigma_{\rm PL}$ cross, 
and the highest final $T_l$ occurs at an intermediate $\sigma_{\rm PL}$ rather than the largest one.

\end{itemize}

The homogeneous evolution for the low-velocity saturation case is shown in Fig.~\ref{fig:saturated_fh_Tl}.
The effects of $\Delta m/m_l$, $m_l$, and $\sigma_{\rm sat}$ follow the similar mechanisms discussed for the power-law case.
The new feature is the saturation velocity $v_0$, which determines the velocity below which the conversion rate saturates.
A smaller $v_0$ delays saturation to later times when $T_l$ is lower, producing a stronger heating effect and preserving a larger heavy fraction.

\section{Linear Structure Formation}
\label{sec:linearForm}

The homogeneous quantities  enter \cref{eq:delta_l,eq:delta_h,eq:theta_l,eq:theta_h} and govern the perturbation evolution through gravitational growth, pressure support, and inelastic conversion. 
We adopt \textit{adiabatic initial conditions}, with the dark sector perturbations initialized following the standard procedure in~\cite{Ma:1995ey}.
As implied by these background quantities, the $s$-wave conversion has negligible impact on structure formation. 
Particularly, the heavy component is depleted before relevant perturbation modes enter the horizon, and the temperature increase of the light component is too small to generate significant pressure support. 
Therefore, the $s$-wave case is not considered further.

\subsection{Evolution of Perturbations}

\begin{figure}
    \centering
    \includegraphics[width=\linewidth]{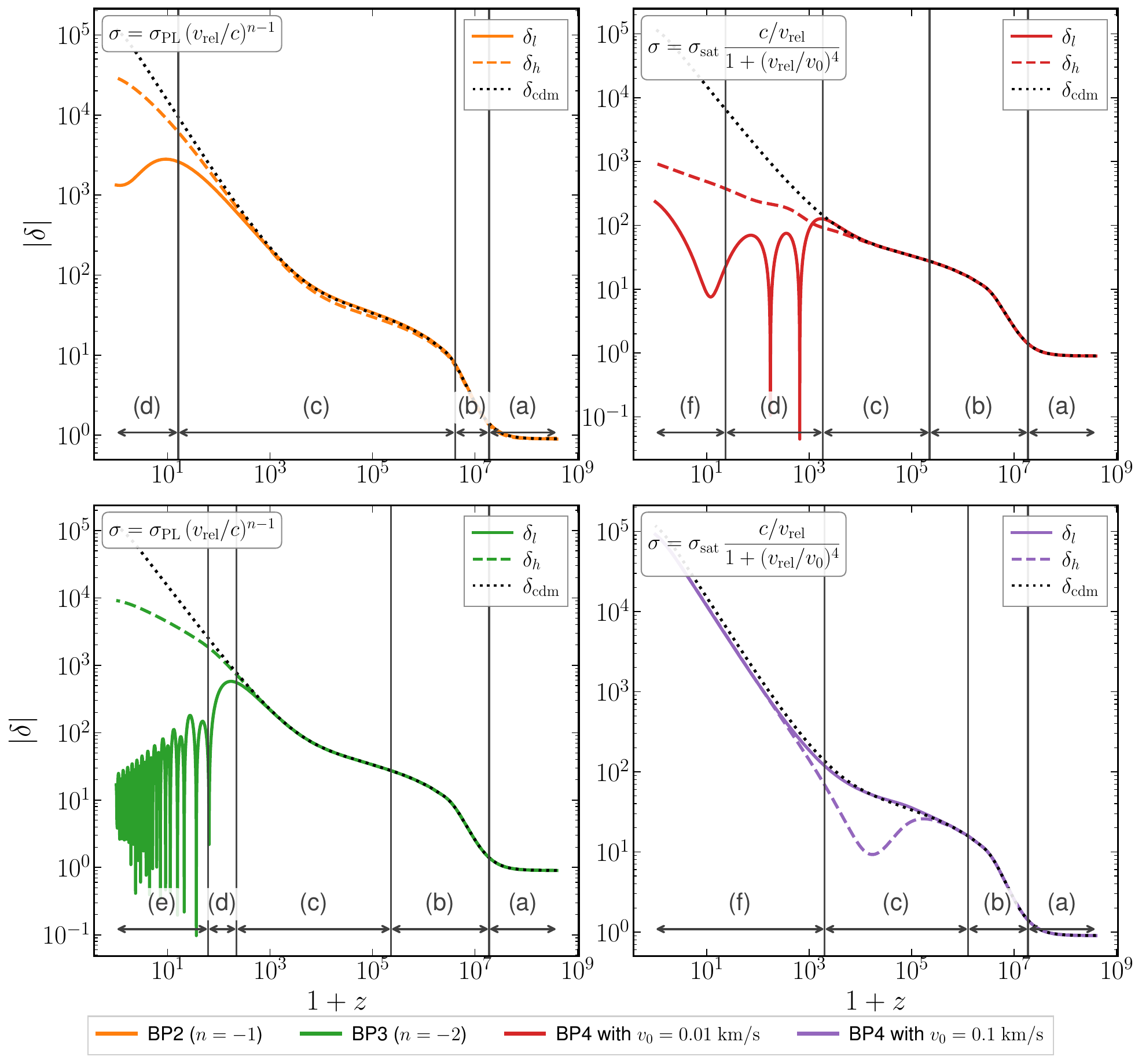}
\caption{
Evolution of the density contrasts for a representative small-scale mode $k = 60\,h\,\mathrm{Mpc}^{-1}$.
Solid and dashed curves show the light and heavy components $\delta_l$ and $|\delta_h|$, respectively. 
The dotted curve is the $\Lambda$CDM reference $|\delta_{\mathrm{cdm}}|$.
The left column uses $\sigma = \sigma_{\mathrm{PL}}(v_{\mathrm{rel}}/c)^{n-1}$ with $n = -1$ (upper left, BP2) and 
$n = -2$ (lower left, BP3), while 
the right column uses $\sigma = \sigma_{\mathrm{sat}}\,(c/v_{\mathrm{rel}})/[1 + (v_{\mathrm{rel}}/v_0)^4]$ with $v_0 = 0.01~\mathrm{km/s}$ (upper right, BP4) and $v_0 = 0.1~\mathrm{km/s}$ (lower right, BP4).
Labeled intervals mark the physical regimes introduced in the text:
\textbf{(a)}~superhorizon,
\textbf{(b)}~early subhorizon growth,
\textbf{(c)}~conversion-driven transition,
\textbf{(d)}~acoustic oscillation,
\textbf{(e)}~acoustic damping, and 
\textbf{(f)}~late-time gravitationally dominated.
}    
    \label{fig:delta_range}
\end{figure}

For $n<0$ and the low-velocity saturation case, 
the later onset of efficient conversion can modify both the density and velocity perturbations. 
By eliminating the velocity divergences from the first-order equations, 
we obtain a set of second-order density perturbation equations (see Appendix~\ref{app:second_order}). 
In these equations, the conversion process introduces additional friction terms $\mathcal{R}_d\,\delta_h'$ and $2\xi\mathcal{R}_d\,\delta_l'$ beyond the standard Hubble drag,
and contributes effective mass terms proportional to $\delta_h$ and $\delta_l$ themselves, such as $(\mathcal{R}_d\mathcal{H}+\mathcal{R}_d')\delta_h$, that directly suppress the overdensity amplitudes.
The source terms $S_h$ and $S_l$ (Eqs.~\eqref{eq:app_Sh} and~\eqref{eq:app_Sl}) encode the gravitational response. 
Additionally, $S_l$ connects to $\delta_h$ that redistributes density between the two components.
To clarify the physical picture, we classify the linear perturbation evolution into six characteristic regimes 
based on 
the conversion rate, the wavenumber, and the Hubble expansion rate.
\begin{itemize}

    \item[\textbf{(a)}] \textbf{Superhorizon Regime:} \\
    In this period, the mode is outside the horizon and satisfies $k/\mathcal{H}\ll 1$.  Causality prevents any modification of the mode on these scales, preserving adiabatic initial conditions. All pressureless components share the same primordial fluctuation, thus both DM species evolve identically to standard CDM.
    
    \item[\textbf{(b)}] \textbf{Early Subhorizon Growth Regime:}\\
    Perturbations begin to evolve causally after entering the horizon ($k/\mathcal{H}\gtrsim 1$). 
    The conversion rate remains much smaller than the Hubble rate.
    Thus all conversion-induced contributions to the perturbation equations are subdominant.
    Both DM species follow the adiabatic cooling and maintain small effective sound speeds. Consequently, $\delta_l$ and $\delta_h$ grow similarly to CDM.

    \item[\textbf{(c)}] \textbf{Conversion-Driven Transition Regime:}\\
    This stage begins when the effects of inelastic conversion first become important at the background level, through either the onset of heavy-component depletion or the departure of $T_l$ from adiabatic cooling. 
    However, the perturbation evolves more gradually. The densities $\delta_l$ and $\delta_h$ initially grow as in CDM, 
    and later deviate from both CDM and each other as the conversion terms become important.

    \item[\textbf{(d)}] \textbf{Acoustic Oscillation Regime:} \\
    Once the pressure term $k^2 \hat{c}_{s,l}^2$ becomes comparable to the gravitational and conversion terms, $\delta_l$ develops acoustic oscillations. 
    The competition between gravitational infall and pressure-restoring forces prevents monotonic growth, forcing $\delta_l$ to oscillate between over- and under-dense states. 
    Therefore, the time-averaged growth rate of perturbations is much lower than the growth rate in a pressureless matter universe. This lower growth rate provides the dominant mechanism behind the suppression of the matter power spectrum on small scales.
    Subsequent damping further reduces the oscillation amplitude.   
    
    \item[\textbf{(e)}] \textbf{Acoustic Damping Regime:} \\
    The friction terms in the light-component evolution equation become cumulatively important
    if the conversion rate remains non-negligible after oscillations begin. 
    These contributions can be characterized by an effective damping rate $\gamma_l \equiv \frac{1}{2}(\mathcal{H} + 2\xi\mathcal{R}_d)$, which includes both Hubble drag and conversion-induced friction. 
    When the damping rate remains smaller than the oscillation frequency, the oscillation amplitude is exponentially suppressed as
    \begin{equation}
        \delta_l \propto \exp\!\left[-\int^{\tau}
        d\tilde{\tau}\,\gamma_l(\tilde{\tau})\right].
    \end{equation}
    Since $\gamma_l$ grows monotonically with the cumulative conversion history, 
    this damping eventually produces a strong suppression of small-scale perturbations.
        
    \item[\textbf{(f)}] \textbf{Late Time Gravitationally Dominated Regime:}\\
    When the light-component temperature returns to adiabatic cooling and all conversion-related terms in the perturbation equations become negligible, the sound speed decreases until $k \hat{c}_{s,l}/\mathcal{H}$ falls below unity and the mode exits the oscillatory regime. 
    Both components then resume standard gravitational growth with $\delta \propto a$ during matter domination.
\end{itemize}

Fig.~\ref{fig:delta_range} illustrates the linear perturbation evolution through the characteristic regimes outlined above, for a representative small-scale mode $k = 60\,h\,\mathrm{Mpc}^{-1}$ using benchmarks BP2, BP3, and BP4. 
Solid and dashed curves show the light and heavy components $|\delta_l|$ and $|\delta_h|$, respectively, with the $\Lambda$CDM reference $|\delta_{\rm cdm}|$ shown as the dotted curve. 
All four panels follow a common sequence of superhorizon evolution \textbf{(a)}, CDM-like subhorizon growth \textbf{(b)}, and a smooth departure from CDM \textbf{(c)} once inelastic conversion becomes dynamically significant. 
The subsequent behavior depends on the competition between the growing cross section at low velocities and the depletion of the heavy component.

For the power-law cases (left column), the evolution beyond regime \textbf{(c)} depends on how rapidly the cross section grows at low velocities.
For BP3 ($n=-2$), the strong velocity enhancement sustains conversion for a sufficient duration to 
drive the mode through acoustic oscillations \textbf{(d)} and significant damping \textbf{(e)}, 
suppressing $|\delta_l|$ by several orders of magnitude relative to $|\delta_{\rm cdm}|$. 
For BP2 ($n=-1$), the weaker enhancement delays the onset of acoustic oscillations to $z \sim 10$, 
resulting in a milder suppression of $|\delta_l|$ than $|\delta_{\rm cdm}|$.

For the low-velocity saturation case (right column), a smaller $v_0$ delays the peak of the conversion rate to a later 
epoch where $\mathcal{H}$ is smaller and the conversion-induced 
heating also produces a larger sound speed . Both effects raise $k \hat{c}_{s,l}/\mathcal{H}$ and determine whether the mode enters the oscillatory regime \textbf{(d)}, as observed for $v_0 = 0.01\,\mathrm{km/s}$ where $|\delta_l|$ develops acoustic oscillations, but not for $v_0 = 0.1\,\mathrm{km/s}$.
In contrast to the power-law cases, the limited duration of efficient conversion 
in the velocity-saturation case means the mode does not enter the damping regime \textbf{(e)}, and 
both cases eventually recover gravitational growth \textbf{(f)} once conversion ceases.

In all scenarios, the heavy component $|\delta_h|$ remains non-oscillatory yet is suppressed relative to CDM. 
This suppression arises directly from the conversion process, 
which slows the growth of $|\delta_h|$ via additional friction $\mathcal{R}_d\,\delta_h'$
and effective mass terms $(\mathcal{R}_d\mathcal{H}+\mathcal{R}_d')\delta_h$, even before light-sector oscillations begin. 
As $|\delta_l|$ is further suppressed, 
the shallower gravitational potential additionally weakens the growth of $|\delta_h|$. 
This correlated suppression of both components leads to a small-scale cutoff in the total matter power spectrum.
\subsection{The Matter Power Spectrum}
\label{sec:pk}
\begin{figure}
    \centering
    \includegraphics[width=\linewidth]{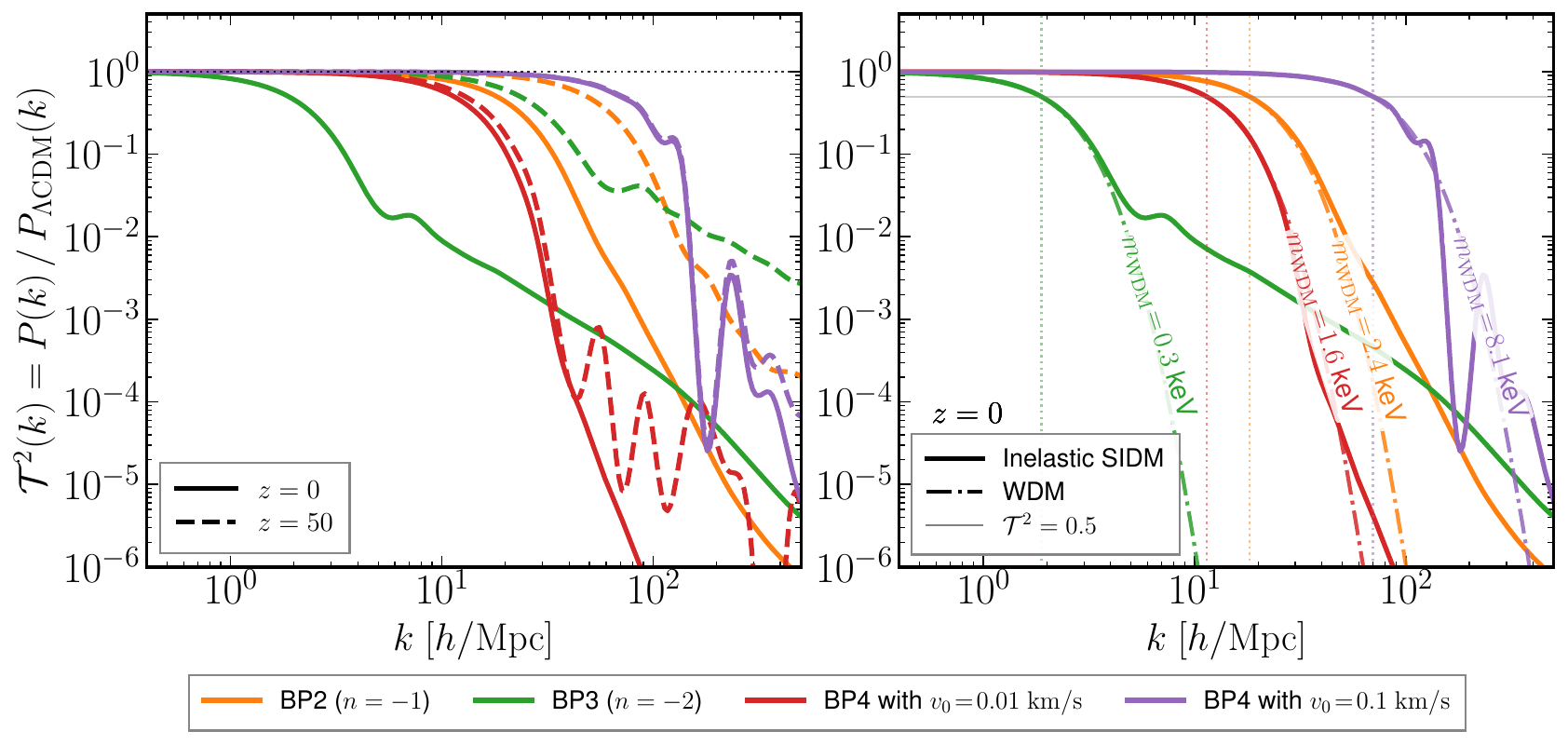}
\caption{Squared transfer function for the benchmark models. 
\textit{Left panel} shows the matter power spectrum ratio $P(k)/P_{\Lambda\mathrm{CDM}}(k)$ at $z=0$ (solid) and $z=50$ (dashed) for four benchmarks $n=-1$ (BP2, orange), $n=-2$ (BP3, green), 
as well as saturated $v_0=0.01\;\mathrm{km\,s^{-1}}$ (BP4, red) and $v_0=0.1\;\mathrm{km\,s^{-1}}$ (BP4, purple). 
\textit{Right panel} compares the $z=0$ transfer functions (solid) to equivalent thermal WDM models (dot‑dashed) matched at 
the half-mode scale $k_{1/2}$ (gray horizontal line), 
with vertical dotted lines marking $k_{1/2}$ for each model and the corresponding equivalent WDM mass indicated.
}
    \label{fig:pk_ratio_wdm}
\end{figure}

Using the Fourier mode evolution established above, 
we now compute the matter power spectrum to capture the scale-dependent impact of inelastic conversion. 
We define the squared transfer function $\mathcal{T}^2(k,z) \equiv P(k,z)/P_{\Lambda\mathrm{CDM}}(k,z)$.
The left panel of Fig.~\ref{fig:pk_ratio_wdm} shows $\mathcal{T}^2(k)$ at $z=0$ (solid) and $z=50$ (dashed) for BP2, BP3, and BP4 (with $v_0 = 0.01$ and $0.1\,\mathrm{km/s}$).
The right panel compares the $z=0$ results with the $\mathcal{T}^2(k)$ of thermal WDM, matched at the half-mode scale $k_{1/2}$ defined by $\mathcal{T}^2(k_{1/2}) = 0.5$.

In the left panel, $n=-2$ shows the strongest suppression, setting in at $k \sim \mathrm{a\,few}\,h\,\mathrm{Mpc}^{-1}$. 
The $n=-1$ case and the low-velocity saturation case with $v_0 = 0.01\,\mathrm{km/s}$ are suppressed starting from $k \sim 10\,h\,\mathrm{Mpc}^{-1}$, while $v_0 = 0.1\,\mathrm{km/s}$ remains close to $\Lambda$CDM until $k \sim 60\,h\,\mathrm{Mpc}^{-1}$. 
Dark acoustic oscillations (DAOs) are visible in both low-velocity saturation cases but absent for the power-law cross sections. 
Comparing $z = 0$ and $z = 50$ for the two velocity-dependence cases, 
the power-law cases show stronger suppression at lower redshift, 
whereas the low-velocity saturation cases show the smooth DAO features from $z = 50$ to $z = 0$.

These features are consistent with the single-mode analysis in the previous section. 
The larger suppression scale for $n=-2$ reflects the prolonged conversion and cumulative damping in regime \textbf{(e)}, which also erases the oscillatory structure 
across modes. 
For $n=-1$, oscillations begin too late to develop appreciable amplitude. 
In low-velocity saturation cases, conversion ends abruptly by $z = 50$, 
fixing the suppression scale as well as cleanly imprinting the pressure-supported phase to produce DAOs.
By contrast, power-law conversion remains active from $z = 50$ to $z = 0$, extending suppression to larger scales.
The subsequent gravitational growth in regime \textbf{(f)} then smooths out the DAO features by $z = 0$.

In the right panel, we show the equivalent WDM masses derived by mapping the half mode scale $k_{1/2}$ of our model to the established analytical relation for a standard thermal relic WDM \cite{Viel:2005qj}. 
These equivalent masses range from $m_{\rm WDM} \simeq 0.3\,\mathrm{keV}$ for $n=-2$ to $8.1\,\mathrm{keV}$ for $v_0 = 0.1\,\mathrm{km/s}$. 
The intermediate cases of $n=-1$ and $v_0 = 0.01\,\mathrm{km/s}$ correspond to $1.6\,\mathrm{keV}$ and $2.4\,\mathrm{keV}$ respectively. 
At high $k$ ($k > k_{1/2}$), the suppression tail in all inelastic SIDM cases is shallower than in their WDM counterparts.
This occurs because the suppression in inelastic SIDM is driven by conversion induced heating rather than simple collisionless free streaming. Furthermore, unlike the smooth cutoff in WDM, the inelastic SIDM transfer functions feature distinct dark acoustic oscillations at high $k$.

\begin{figure}[t]
    \centering
    \includegraphics[width=\textwidth]{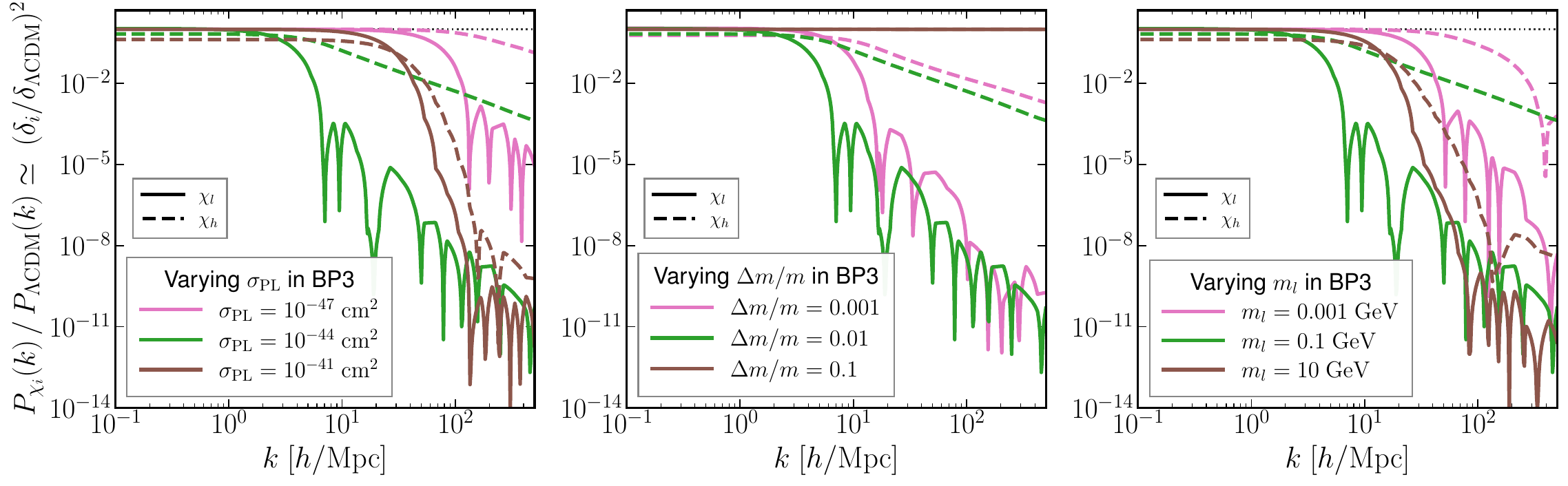}
    \caption{Parameter dependence of the linear power spectra for $\sigma = \sigma_{\rm PL}(v_{\rm rel}/c)^{-3}$, shown as ratios to the $\Lambda$CDM matter power spectrum at $z=0$. 
    The contribution of each component is approximated by $P_{\chi_i}(k) \simeq P_{\Lambda\mathrm{CDM}}(k)\,[\delta_i(k)/\delta_{\Lambda\mathrm{CDM}}(k)]^2$, with the light component $\chi_l$ in solid lines and the heavy component $\chi_h$ in dashed lines. 
    In each panel, one parameter is varied around the BP3 fiducial values (green curves: $\sigma_{\rm PL} = 10^{-44}\;\mathrm{cm^2}$, $\Delta m/m = 0.01$, $m_l = 0.1\;\mathrm{GeV}$), with the other two fixed.
}
    \label{fig:pk_ratio_lh}
\end{figure}

Fig.~\ref{fig:pk_ratio_lh} shows the power spectra of the light (solid lines) and heavy (dashed lines) components separately, fixing the power-law cross section at $n = -2$ and varying $\sigma_{\rm PL}$ (left), $\Delta m / m$ (center), or $m_l$ (right) around the BP3 fiducial values. Both components remain close to unity on large scales. However, on smaller scales, the light component exhibits an earlier and sharper turnover with pronounced DAO features, due to direct pressure support from conversion-induced heating. In contrast, the heavy component shows a smoother suppression governed by its depletion history and the weakened gravitational potential.

A notable feature is the non-monotonic evolution of the power spectra. 
The acoustic suppression is present, and the cutoff shifts to the largest scales, 
at the intermediate BP3 fiducial values (green curves). 
Any deviation from the reference values of $\sigma_{\rm PL}$, $\Delta m / m$, or $m_l$ weakens the suppression and 
shifts the cutoff to smaller scales.
This behavior reflects the combined effects of the total energy available for conversion, 
the timing and duration of the heating episode, and the persistence of the resulting pressure support.
A very small $\sigma_{\rm PL}$ results in inefficient conversion, 
while an excessively large value causes early depletion of the heavy species and shortens the effective heating period. 
Similarly, a large $\Delta m / m$ reduces the initial heavy fraction in thermal equilibrium. 
Varying $m_l$ shifts both the onset redshift and the duration of rapid abundance evolution. 
In all cases, the heavy component is suppressed along with the light component, 
confirming that gravitational coupling efficiently transfers acoustic suppression from the light sector to the pressureless heavy sector.

\section{Constraints}
\label{sec:constraints}
In this section we derive constraints on the parameter space using small-scale structure probes. 
The Lyman-$\alpha$ forest and high-redshift UV luminosity function constraints are taken from 
existing works~\cite{Murgia:2018now,Sabti:2021bff} 
that calibrated their limits against nonlinear simulations but projected them onto the linear matter power spectrum. 
We apply these recast constraints directly to our linear power spectra. 
Sec.~\ref{sec:lya_recast} describes the Lyman-$\alpha$ recast, 
Sec.~\ref{sec:uvlf_recast} presents the UV luminosity function recast, and 
Sec.~\ref{sec:paramter_space} combines the exclusions to identify the allowed parameter space.

\subsection{\texorpdfstring{Lyman-$\alpha$ forest}{Lyman-alpha forest}}
\label{sec:lya_recast}

The Lyman-$\alpha$ forest consists of neutral-hydrogen absorption lines in quasar spectra at $z\simeq2$--$5$ and 
is one of the most stringent probes of small-scale matter clustering. 
State-of-the-art analyses of high-resolution data from XQ-100~\cite{Lopez:2016bfc} and HIRES/MIKE~\cite{Viel:2013apy} have placed strong lower bounds on the thermal WDM mass~\cite{Irsic:2017ixq}: 
$m_{\rm WDM} > 5.3\,\mathrm{keV}$ at 95\% C.L. from a full Markov chain Monte Carlo analysis comparing hydrodynamical simulations with observational data.

In this work, we cannot repeat such computationally expensive hydrodynamical simulations for every point in our parameter space,  
thus we adopt the recast approach validated by Ref.~\cite{Murgia:2018now}. 
The central premise is that if a model beyond $\Lambda$CDM produces a linear matter power spectrum with cumulative suppression comparable to that of a $5.3\,\mathrm{keV}$ thermal WDM, 
it will yield statistically indistinguishable Lyman-$\alpha$ fluxes. 
This approach is particularly well suited to our cases, 
whose transfer function shape only differs from that of WDM due to the oscillatory features and shallower asymptotic slope discussed 
in the right panel of Fig.~\ref{fig:pk_ratio_wdm}.

Following Ref.~\cite{Murgia:2018now}, we first compute the one-dimensional matter power spectrum,
\begin{equation}
    P_{1\mathrm{D}}(k)=\frac{1}{2\pi}\int_{k}^{\infty}\mathrm{d}k'\,k'\,P(k'),
\label{eq:P1D_def}
\end{equation}
and define the suppression estimator
\begin{equation}
    A \equiv \int_{k_{\min}}^{k_{\max}}\mathrm{d}k\,
    \frac{P_{1\mathrm{D}}^{\rm model}(k)}{P_{1\mathrm{D}}^{\Lambda{\rm CDM}}(k)},
    \label{eq:A_def}
\end{equation}
with the fractional deficit
\begin{equation}
    \delta A \equiv \frac{A_{\Lambda{\rm CDM}}-A}{A_{\Lambda{\rm CDM}}}.
    \label{eq:deltaA_def}
\end{equation}
We adopt $k_{\min} = 0.5\;h\,\mathrm{Mpc}^{-1}$ and $k_{\max} = 20\;h\,\mathrm{Mpc}^{-1}$ to cover the wavenumber range probed by the XQ-100 and HIRES/MIKE data~\cite{Irsic:2017ixq,Murgia:2018now}. The exclusion threshold is calibrated by evaluating $\delta A$ for a thermal relic with $m_{\rm WDM}^{\rm ref} = 5.3\,\mathrm{keV}$, yielding
\begin{equation}
    \delta A_{\rm crit} \equiv \delta A(m_{\rm WDM}^{\rm ref}) \simeq 0.31.
    \label{eq:deltaA_crit}
\end{equation}

The exclusion condition is as follows: any parameter point with $\delta A > \delta A_{\rm crit}$ is excluded at the $2\sigma$ level. 
We verify that the exclusion region obtained from matching the half-mode scale $k_{1/2}$ (where $\mathcal{T}^2 = 0.5$) is almost identical to this condition for all cross-sections. 
Therefore, we present only the exclusion derived from this condition.

\subsection{\texorpdfstring{High-$z$ UV luminosity functions}{High-z UV luminosity functions}}
\label{sec:uvlf_recast}

High-redshift UV galaxy luminosity functions (UVLFs) offer an independent probe of small-scale matter clustering during cosmic dawn and reionization. 
UV-selected galaxies at $z\sim4$–$10$ reside in low-mass halos. 
Their abundance as a function of UV magnitude encodes the halo mass function and 
indirectly the linear matter power spectrum on comoving scales that are otherwise difficult to access.

We adopt the approach of Ref.~\cite{Sabti:2021bff}, which uses current HST UVLF measurements to infer the small-scale matter clustering amplitude after marginalising over astrophysical parameters. 
Rather than performing a full UVLF likelihood analysis, 
we recast the model-independent power-spectrum measurements reported therein as constraints on the transfer function $\mathcal{T}^2(k)$. 
Following Ref.~\cite{Sabti:2021bff}, we divide the transfer function into two $k$ bins. 
The measured amplitudes ${a_{s,i}}$ relative to $\Lambda$CDM are:
$a_{s,2} = 0.93^{+0.34}_{-0.25}$ for $0.5~\mathrm{Mpc}^{-1} \leq k < 2.25~\mathrm{Mpc}^{-1}$ and
$a_{s,3} = 0.66^{+0.43}_{-0.17}$ for $2.25~\mathrm{Mpc}^{-1} \leq k < 10~\mathrm{Mpc}^{-1}$,
both at 68\% CL after marginalisation over all cosmological and astrophysical parameters.

To project our theoretical models onto these constraints, we compute the bin-averaged transfer function to estimate the effective amplitude for each point in our parameter space:

\begin{equation}
    \overline{\mathcal{T}^2_i}
    \;\equiv\;
    \frac{1}{k_{\max,i}-k_{\min,i}}
    \int_{k_{\min,i}}^{k_{\max,i}} \mathrm{d}k \; \mathcal{T}^2(k)\,.
    \label{eq:T2bar_uvlf}
\end{equation}

We then evaluate this theoretical expectation against the measured posteriors of $a_{s,i}$ in Ref.~\cite{Sabti:2021bff}. Adopting a conservative recasting approach, we exclude any parameter where the predicted $\overline{\mathcal{T}^2_i}$ falls below the $2\sigma$ lower bound of the measured amplitude in either bin (i.e.,$\overline{\mathcal T_i^2} < a_{s,i} - 2|\sigma_{-,i}|$, where $\sigma_{-,i}$ is the lower uncertainty for $a_{s,i}$).

We would like to emphasise that this UVLF procedure is a recast-level constraint rather than a full UVLF likelihood analysis, 
which would require an explicit halo model, a galaxy--halo connection, and a consistent marginalisation over astrophysical parameters. 
Our goal here is to 
simply examine a given parameter point producing small-scale suppression generically incompatible with 
current UVLF-inferred clustering. 
A dedicated likelihood analysis is left to future work.

\subsection{Allowed parameter space}
\label{sec:paramter_space}
\begin{figure}
    \centering
    \includegraphics[width=\linewidth]{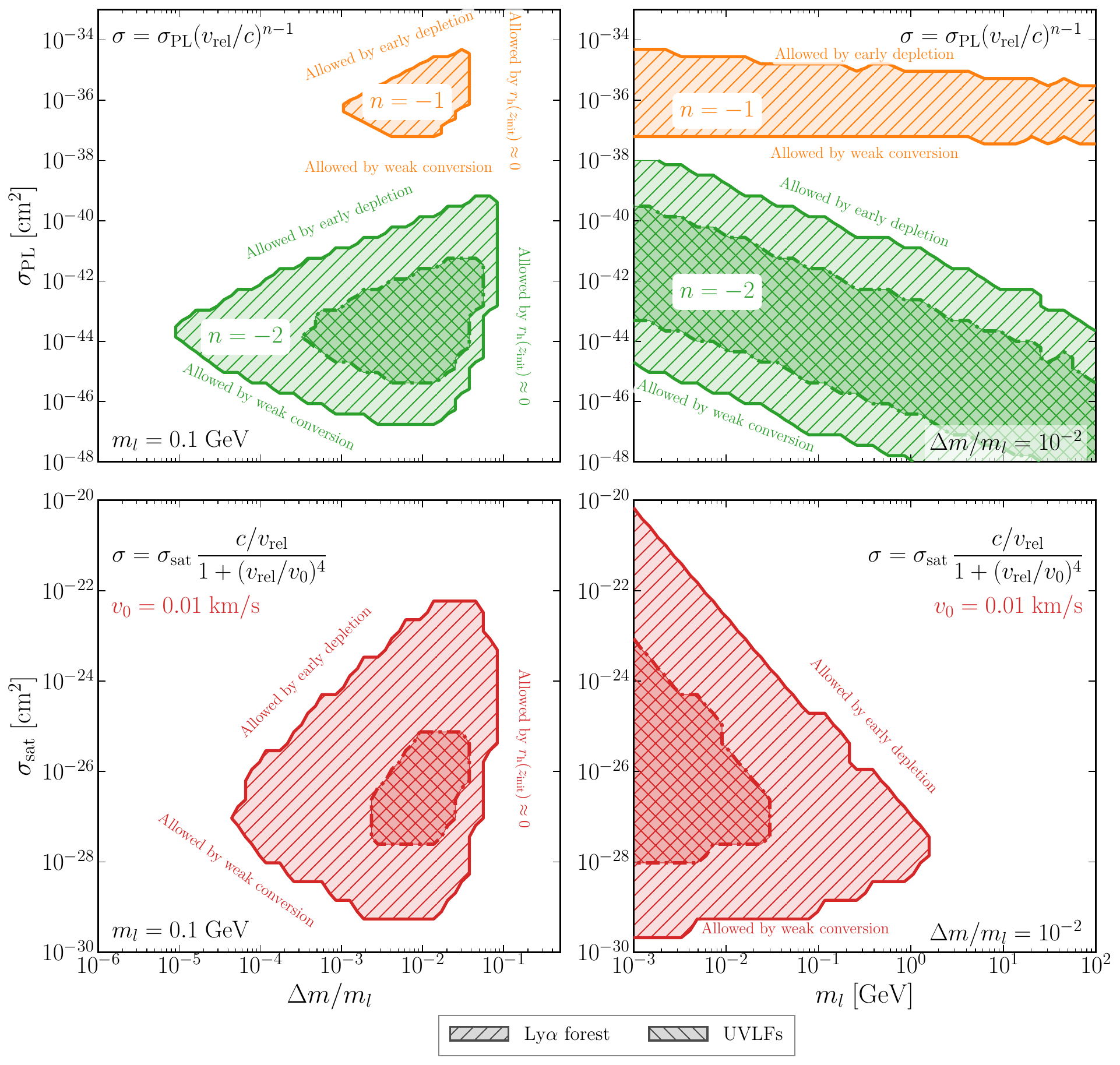}
    \caption{Excluded regions in the $(\Delta m/m_l,\,\sigma_{\rm PL}$ and $\sigma_{\rm sat})$ plane
    (left column, $m_l = 0.1\;\mathrm{GeV}$) and the
    $(m_l,\,\sigma_{\rm PL}$ and $\sigma_{\rm sat})$ plane (right column, $\Delta m/m_l = 10^{-2}$)
    from the Ly-$\alpha$ forest area criterion ($/$-hatched)
    and high-redshift UVLFs ($\backslash$-hatched).
    \textit{Top row:} power-law cross section with $n=-2$
    (green) and $n=-1$ (orange).
    \textit{Bottom row:} Low-velocity saturation cross section
    with $v_0 = 0.01\;\mathrm{km\,s^{-1}}$ (red).}
    \label{fig:constraints_combined}
\end{figure}

Fig.~\ref{fig:constraints_combined} presents the exclusion regions from the Lyman-$\alpha$ (outer contour) 
and high-redshift UVLFs (inner contour).
The upper panels show the power-law scenarios ($n=-1$ and $n=-2$), and the lower panels show the low-velocity saturation scenarios.
In the left panels ($\Delta m/m_l$ vs. $\sigma_{\rm PL}$ and $\sigma_{\rm sat}$), $m_l$ is fixed to $0.1~\mathrm{GeV}$. 
In the right panels ($m_l$ vs. $\sigma_{\rm PL}$ and $\sigma_{\rm sat}$), we adopt $\Delta m/m_l = 10^{-2}$.

As discussed in Sec.~\ref{sec:pk}, the excluded regions form closed contours, 
occurring within an intermediate range of cross-section normalization. 
Below this range, exothermic conversion is too weak to impact perturbations, 
while above it, the heavy DM species depletes so rapidly that heating ceases. 
Within certain cross-section, exothermic conversion generates significant pressure support on scales probed 
by the Lyman-$\alpha$ forest and UVLFs. 
A similar non-monotonic behavior appears along the $\Delta m/m_l$ axis. 
At large splittings, exclusion is limited by the exponential suppression of the initial heavy fraction, 
whereas at small splittings it is constrained by insufficient energy release per scattering.
Clearly, constraints from UVLFs are weaker than those from Lyman-$\alpha$ because they probe larger scales.
For the power-law scenario (two upper panels of Fig.~\ref{fig:constraints_combined}), 
the excluded region spans $\sigma_{\rm PL} \sim 10^{-48}$ to $10^{-40}~\mathrm{cm}^2$ 
in the $n=-2$ scenario and $\sigma_{\rm PL} \sim 10^{-37}$ to $10^{-35}~\mathrm{cm}^2$ in the $n=-1$ scenario. 
The $n=-2$ scenario excludes a wider interval of cross sections, 
owing to its stronger low-velocity enhancement that sustains the conversion process to later times. 
In contrast, the low-velocity saturation scenario (two lower panels of Fig.~\ref{fig:constraints_combined}) 
shifts the excluded region to larger cross-section, roughly $\sigma_{\rm sat} \sim 10^{-30}$ to $10^{-22}~\mathrm{cm}^2$, 
because efficient energy conversion is confined to a specific velocity window, 
requiring a larger cross-section to achieve sufficient total energy injection.

\begin{figure}
    \centering
    \includegraphics[width=\linewidth]{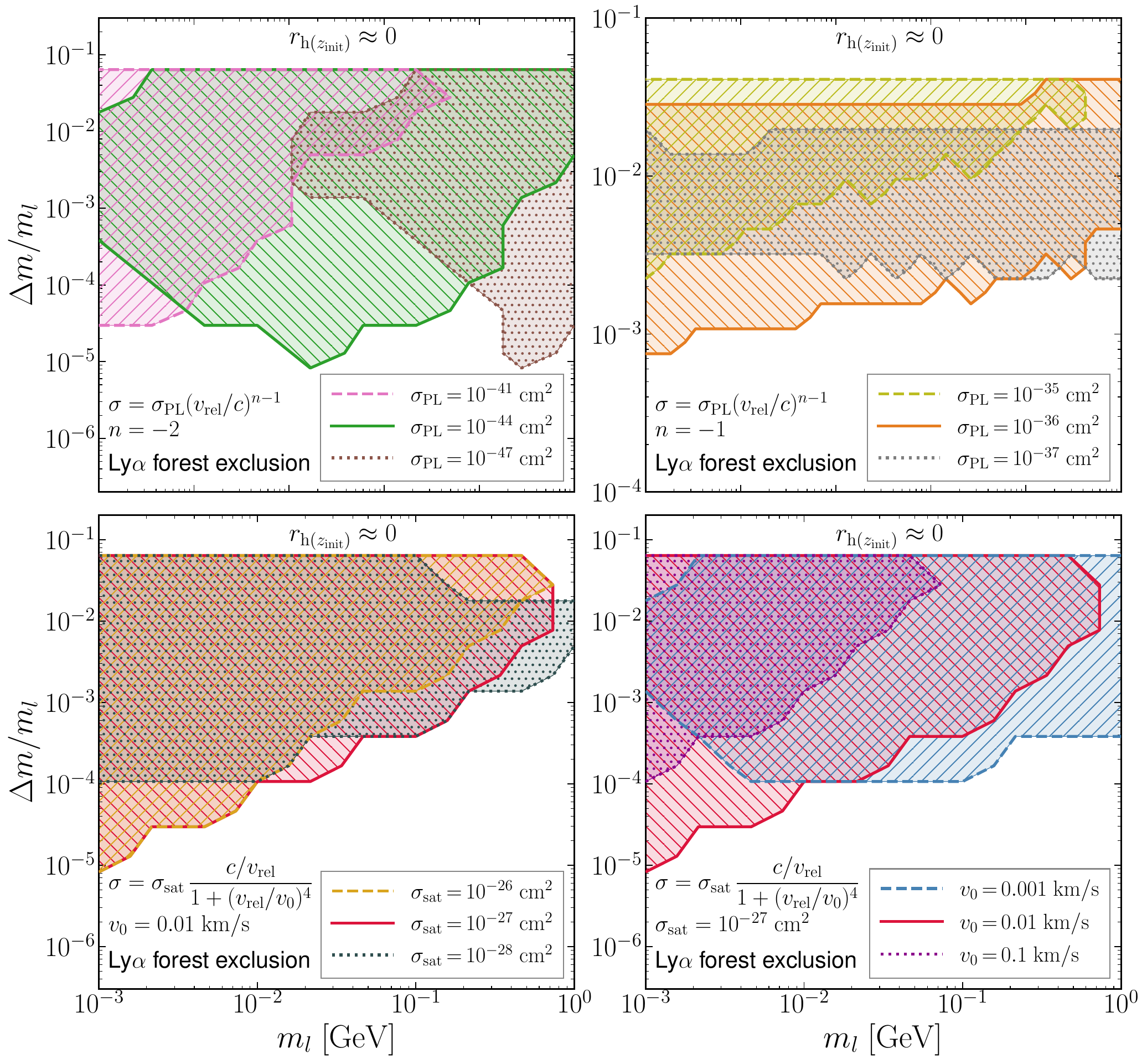}
    \caption{Ly-$\alpha$ forest exclusion in the
    $(m_l,\,\Delta m/m_l)$ plane.
    \textit{Top Left:} Power-law cross section with $n=-2$
    ($\sigma_{\rm PL} = 10^{-41},\,10^{-44},\,10^{-47}\;\mathrm{cm}^2$).
    \textit{Top Right:} Power-law cross section with $n=-1$
    ($\sigma_{\rm PL} = 10^{-35},\,10^{-36},\,10^{-37}\;\mathrm{cm}^2$).
    \textit{Bottom Left:} Low-velocity saturation cross section with fixed $v_0 = 0.01\;\mathrm{km\,s^{-1}}$
    ($\sigma_{\rm sat} = 10^{-26},\,10^{-27},\,10^{-28}\;\mathrm{cm}^2$).
    \textit{Bottom Right:} Low-velocity saturation cross section with fixed $\sigma_{\rm sat} = 10^{-27}\;\mathrm{cm}^2$ for three transition velocity values 
    ($v_0 = 0.001,\,0.01,\,0.1\;\mathrm{km\,s^{-1}}$).}
    \label{fig:constraints_mass}
\end{figure}

Fig.~\ref{fig:constraints_mass} shows the Lyman-$\alpha$ exclusion in the $(m_l,\,\Delta m/m_l)$ plane for representative cross sections, namely $\sigma_{\rm PL}$ (upper panels) and $\sigma_{\rm sat}$ (lower panels). 
Larger cross sections shift the excluded region toward smaller $\Delta m/m_l$ and larger $m_l$, as a higher conversion rate compensates for a smaller energy release or a heavier $m_l$.
However, once the cross-section becomes too large as seen in Fig.~\ref{fig:constraints_combined}, the constraining power disappears. 
The upper boundary of all velocity-dependent cross-sections is determined by the $r_h(z_{\rm init}) \approx 0$ condition, 
due to the Boltzmann suppression of the initial heavy fraction. 
For cross sections with steep velocity dependence ($n=-2$ and the low-velocity saturation case), 
low-velocity enhancements amplify the exothermic conversion, leading to nearly identical bounds. 
In the saturated parametrization, the interaction rate scales as $(v_0/v_{\rm rel})^4$ when $v_0 \ll v_{\rm rel}$. 
Among the three choices of $v_0$, the case $v_0 = 0.01~\mathrm{km~s^{-1}}$ (red solid line) with $\sigma_{\rm sat}=10^{-27}~\mathrm{cm}^2$ yields the widest redshift window for efficient conversion, whereas the exclusion regions shrink for $v_0 = 0.001~\mathrm{km~s^{-1}}$ (blue dashed line) and $v_0 = 0.1~\mathrm{km~s^{-1}}$ (purple dotted line) due to suppression from the $v_0^4$ factor and the condition $v_0 \approx v_{\rm rel}$, respectively. 
Unlike these two scenarios, the weaker velocity dependence of the $n=-1$ case causes an earlier drop in exothermic conversion, making its exclusions change with respect to $\sigma_{\rm PL}$.

In summary, cosmological probes like Lyman-$\alpha$ can constrain certain mass splittings independently of 
other DM detection methods that rely on interactions with visible particles or dark radiation. 
In our default setup (inelastic SIDM with thermal initial conditions) and 
for the most optimistic cases (solid lines) at $m_l\approx 100~\mathrm{MeV}$, 
current Lyman‑$\alpha$ data can limit $\Delta m/m_l$ 
to $\mathcal{O}(10^{-5})$ and $\mathcal{O}(10^{-3})$ for $n=-2$ and $n=-1$, respectively. 
For the low-velocity saturation scenarios, the exclusion depends strongly on $m_l$ and $v_0$. 
With the optimistic parameters $\sigma_{\rm sat}=10^{-27}~\mathrm{cm}^2$ and $v_0 = 0.01~\mathrm{km~s^{-1}}$, 
the largest allowed $\Delta m/m_l$ is $\mathcal{O}(10^{-5})$ for $m_l=1~\mathrm{MeV}$ and 
$\mathcal{O}(10^{-4})$ for $m_l=10~\mathrm{MeV}$.

\section{Summary and Discussion}
\label{sec:summary}
In this work, we have investigated the linear cosmological evolution of inelastic SIDM in a two-component dark sector with a small mass splitting $\Delta m$. 
We derived the coupled background and linear perturbation equations for the conversion process $\chi_h\chi_h \leftrightarrow \chi_l\chi_l$, considering both power-law and low-velocity saturation cross sections and 
using a modified Boltzmann solver based on \texttt{CLASS}.

The homogeneous and perturbation evolution are shaped by the interplay of four parameters. The velocity index $n$ (or saturation velocity $v_0$) controls the epoch and duration of efficient conversion. The mass splitting $\Delta m/m_l$ plays a dual role, setting the energy release per conversion while simultaneously suppressing the initial heavy fraction via the Boltzmann factor. The DM mass $m_l$ sets $z_{\rm init}$ and $T_{\rm init}$, and adjusts departure times of the conversion history. The cross-section normalization governs the overall conversion strength, but an excessively large value causes premature depletion of the heavy species. These competing effects produce non-monotonic parameter dependence throughout. At the perturbation level, conversion-induced heating generates pressure support in the light component. This pressure support, together with conversion-induced drag between the two species, drives acoustic oscillations and suppresses small-scale density growth. The heavy component, though pressureless, is suppressed concurrently through additional friction terms and the weakened gravitational potential. The resulting transfer function features a cutoff at $k \gtrsim 1\;h\,\mathrm{Mpc}^{-1}$ accompanied by DAOs. The visibility of DAOs depends on the cross-section type. Low-velocity saturation scenarios, where conversion terminates abruptly, preserve distinct oscillatory features, whereas power-law cross sections with sustained late-time conversion progressively smooth them out. Compared to thermal WDM matched at the same half-mode scale, the inelastic SIDM transfer function exhibits a shallower high-$k$ suppression tail and the presence of DAOs. The shallower tail reflects the fact that the suppression originates from conversion-induced heating rather than collisionless free streaming. These distinct spectral features offer potential 
model discrimination. We summarize the model parameters and their dependency in Table~\ref{tab:model_parameters}.

Using recast constraints from Lyman-$\alpha$ forest data and high-redshift UV luminosity functions, we identified closed exclusion regions driven by the non-monotonic parameter dependences. For the power-law parametrization, the excluded range spans $\sigma_\mathrm{PL} \sim 10^{-48}$--$10^{-40}\;\mathrm{cm}^2$ for $n=-2$ and $\sim 10^{-37}$--$10^{-35}\;\mathrm{cm}^2$ for $n=-1$, probing mass splittings down to $\mathcal{O}(10^{-5})$ and $\mathcal{O}(10^{-3})$, respectively. The low-velocity saturation scenario shifts the exclusion to $\sigma_\mathrm{sat} \sim 10^{-30}$--$10^{-22}\;\mathrm{cm}^2$. These results demonstrate that the internal thermodynamics of a secluded dark sector can be constrained purely through its gravitational imprint on structure formation, independent of DM--SM couplings.

Future extensions include $N$-body simulations with inelastic conversion to capture nonlinear evolution and halo formation, exploration of non-thermal initial conditions, and comprehensive Markov chain Monte Carlo analyses to refine the allowed parameter space.
Next generation probes such as 21~cm cosmology and CMB spectral distortion measurements will further extend the testable parameter space.

\begin{table}[t]
\centering
\caption{Summary of model parameters.}
\label{tab:model_parameters}
\begin{tabular}{lll}
\hline
Parameter & Definition & Physical role \\
\hline
$m_l$ & Light-state mass & Sets the mass and temperature scale. \\
$\Delta m/m_l$ & Fractional mass splitting & Controls energy release and initial abundance. \\
$\sigma_{\rm PL}$ & Cross-section normalization & Sets the conversion strength. \\
$n$ & Velocity index & Controls velocity dependence. \\
$\sigma_{\rm sat}$ & Cross-section normalization & Sets the cross-section with $v_{\rm rel}\to 0$. \\
$v_0$ & Saturation velocity & Sets the saturation epoch. \\
\hline
\end{tabular}
\end{table}

\section*{Acknowledgments}
We would like to thank Peizhi Du, Yi Wang and Yi-Ming Zhong for insightful discussions.
This work is supported in part by 
the National Natural Science Foundation of China (No. 12588101), 
the National Key Research and Development Program of China (No. 2022YFF0503304), 
the Project for Young Scientists in Basic Research of the Chinese Academy of Sciences (No. YSBR-092), and 
the China Manned Space Program (No. CMS-CSST-2025-A03), the Basic and Frontier Research Project of PCL (grant No. 2025QYB012) and the Major Key project of Peng Cheng Laboratory.

\newpage
\appendix
\section{Appendix}

\subsection{Thermally averaged conversion rates}
\label{app:rates}

For a Maxwell--Boltzmann distribution with temperature $T$,
the thermally averaged product $\langle \sigma v_{\rm rel} \rangle$ 
is defined as
\begin{equation}
    \langle \sigma v_{\rm rel} \rangle
    = \frac{1}{2}\left(\frac{m}{4\pi T}\right)^{3/2}
    \int_{v_{\rm min}}^{\infty} \sigma\, v_{\rm rel}\,
    e^{-m v_{\rm rel}^2/4T}\,
    4\pi v_{\rm rel}^2\, dv_{\rm rel},
    \label{eq:thermal_avg}
\end{equation}
where $m/2$ is the reduced mass of identical particles
and $v_{\rm min} = 0$ for the exothermic channel, 
$v_{\rm min} = 2\sqrt{2\Delta m/m_l}$ for the endothermic channel.
Substituting the cross-section parameterizations of 
Sec.~\ref{sec:models}, the conversion rates take the 
following explicit forms.

\begin{itemize}

\item \textbf{Power-law parameterization.}
Inserting $\sigma = \sigma_{\rm PL}\,v_{\rm rel}^{\,n-1}$
into Eq.~\eqref{eq:thermal_avg}, and using $\mathcal{R}\equiv a \frac{\rho}{m} \langle \sigma v_{\rm rel} \rangle$,
the exothermic and endothermic conversion rates are
\begin{align}
    \mathcal{R}_d^{\rm PL}
    &= a\,\frac{\bar\rho_h}{m_h}\,
    c_n\,\sigma_{\rm PL}
    \left(\frac{T_h}{m_h}\right)^{n/2},
    \label{eq:Rd_PL} \\[6pt]
    \mathcal{R}_u^{\rm PL}
    &= a\,\frac{\bar\rho_l}{m_l}\,
    c_n\,\sigma_{\rm PL}
    \left(\frac{T_l}{m_l}\right)^{n/2}\,
    Q\!\left(\tfrac{3}{2}+\tfrac{n}{2},\;
    \frac{2\Delta m}{T_l}\right),
    \label{eq:Ru_PL}
\end{align}
where
\begin{equation}
    c_n \equiv \frac{2^{n}\,
    \Gamma\!\left(\tfrac{3}{2}+\tfrac{n}{2}\right)}{\sqrt\pi}
\end{equation}
is a numerical prefactor from the thermal average, and $Q(x,y) \equiv \Gamma(x,y)/\Gamma(x)$ is the regularised upper incomplete gamma function. 
The factor $Q$ encodes the kinematic threshold for the endothermic process: $Q \to 1$ when $\Delta m/T_l \to 0$ (threshold irrelevant) and $Q \to 0$ when $\Delta m/T_l \gg 1$ (Boltzmann-suppressed).
In the latter limit, $\mathcal{R}_u$ reduces to $\mathcal{R}_d\,e^{-2\Delta m/T_l}$ up to a power-law prefactor.

\item \textbf{Low-velocity saturation.}
Inserting Eq.~\eqref{eq:sigma_saturated} into 
Eq.~\eqref{eq:thermal_avg}, the exothermic rate is
\begin{equation}
    \mathcal{R}_d^{\rm sat}
    = a\,\frac{\bar\rho_h}{m_h}\,
    \sigma_{\rm sat}\,
    \left[1 + \left(\frac{3\sqrt\pi}{16}\right)^{4/3}
    \left(\frac{4\,T_h}{m_h\,v_0^2}\right)^{2}
    \right]^{-3/4},
    \label{eq:Rd_sat}
\end{equation}
and the endothermic rate is approximated by
\begin{equation}
    \mathcal{R}_u^{\rm sat}
    \simeq \mathcal{R}_d^{\rm sat}\;
    e^{-2\Delta m/T_l}.
    \label{eq:Ru_sat}
\end{equation}
This approximation is appropriate because the saturation cross-section targets the low-velocity regime where
$\Delta m/T_l \gtrsim 1$; the kinematic threshold then lies deep in the Boltzmann tail, and the exponential
factor captures the dominant suppression.

\end{itemize}
\subsection{Evolution of Number Density}
\label{sec:num_density}

We consider a dark sector consisting of two non-relativistic species, denoted as $\chi_l$ (light) and $\chi_h$ (heavy), which interconvert via the inelastic scattering process:
\begin{equation}
    \chi_l\chi_l \;\leftrightarrow\; \chi_h\chi_h \,.
\end{equation}
and assume that each species maintains an approximate Maxwell--Boltzmann distribution characterized by its own temperature $T_i$ (allowing for $T_l \neq T_h$). 
For the collision term, we account for the symmetry factor of $1/2$ for identical initial states and the fact that each reaction event changes the particle number by two. These factors cancel out, allowing the net source terms to be written compactly as $\pm\langle\sigma v\rangle n_i^2$.

In an expanding Friedmann-Robertson-Walker (FRW) background, the Boltzmann equations for the number densities are given by:
\begin{align}
    \dot{n}_l + 3Hn_l
    &= -\langle\sigma v\rangle_{ll\to hh}(T_l)\,n_l^2
    + \langle\sigma v\rangle_{hh\to ll}(T_h)\,n_h^2 \,,
    \label{eq:nl_dot}\\
    \dot{n}_h + 3Hn_h
    &= -\langle\sigma v\rangle_{hh\to ll}(T_h)\,n_h^2
    + \langle\sigma v\rangle_{ll\to hh}(T_l)\,n_l^2 \,.
    \label{eq:nh_dot}
\end{align}
For convenience, we define the total number density $n_{\rm tot}$ and the heavy-state fraction $r_h$ as:
\begin{equation}
    n_{\rm tot} \equiv n_l + n_h, \qquad 
    r_h \equiv \frac{n_h}{n_{\rm tot}} \,.
    \label{eq:fh_def}
\end{equation}
Since the total number density scales as $n_{\rm tot} \propto a^{-3}$ (implying $\dot{n}_{\rm tot} + 3H n_{\rm tot} = 0$), the evolution of the heavy fraction is driven solely by the interaction terms. Combining Eqs.~\eqref{eq:nl_dot}--\eqref{eq:fh_def}, we obtain:
\begin{equation}
    \dot{r}_h = n_l T_l \langle\sigma v\rangle_{ll\to hh}\times \left(1-r_h\right) 
              - n_h T_h \langle\sigma v\rangle_{hh\to ll}\times r_h \,.
    \label{eq:fh_t}
\end{equation}
It is often convenient to rewrite the evolution in terms of redshift $z$. Using the relation $dt/dz = -[ (1+z)H(z) ]^{-1}$, Eq.~\eqref{eq:fh_t} becomes:
\begin{equation}
    \frac{\mathrm{d} r_h}{\mathrm{d} z} = \frac{1}{(1+z)H(z)} 
    \left[ 
        n_h \langle\sigma v\rangle_{hh\to ll}(T_h) r_h  
        - n_l \langle\sigma v\rangle_{ll\to hh}(T_l)(1-r_h) 
    \right] \,.
    \label{eq:fh_z}
\end{equation}

\subsection{Evolution of Temperature}
\label{sec:temp_evolution}

We model each species as a non-relativistic ideal gas with energy density $\rho_i$ and pressure $p_i$:
\begin{equation}
    \rho_i = m_i n_i + \frac{3}{2}n_i T_i, \qquad p_i = n_i T_i \,.
    \label{eq:rho_p}
\end{equation}
Let $U_i \equiv \rho_i V$ be the internal energy in a comoving volume $V$. According to the first law of thermodynamics,
\begin{equation}
    \mathrm{d}U_i = -p_i\,\mathrm{d}V + \mathrm{d}Q_i \,,
    \label{eq:firstlaw}
\end{equation}
where $\mathrm{d}Q_i$ represents the heat exchange due to the inelastic conversion.

Over a time interval $\mathrm{d}t$, the net energy transferred to the
light species $\chi_l$ is
\begin{equation}
    \frac{\mathrm{d}Q_l}{\mathrm{d}t}
    =
    V\left[
        \langle\sigma v\rangle_{hh\to ll}(T_h)\,n_h^2
        \left(m_h+\epsilon_{{\rm rem},h}\right)
        -
        \langle\sigma v\rangle_{ll\to hh}(T_l)\,n_l^2
        \left(m_l+\epsilon_{{\rm rem},l}\right)
    \right],
    \label{eq:Ql_dot}
\end{equation}
where $\epsilon_{{\rm rem},i}$ is the reaction-rate-weighted mean kinetic
energy carried by each particle removed from the source population
$\chi_i$.

When the reaction kernel $\sigma_i v_{\rm rel}$ is independent of the
relative velocity, the reaction does not preferentially select particles
of a particular kinetic energy, and   $\epsilon_{{\rm rem},i}$ reduces to
the ordinary thermal average,
\begin{equation}
    \epsilon_{{\rm rem},i}=\frac{3}{2}T_i.
\end{equation}
For a velocity-dependent kernel, the rate weighting produces a correction
of order $T_i$. For example, if
$\sigma_i v_{\rm rel}\propto v_{\rm rel}^{\,n}$, one obtains
\begin{equation}
    \epsilon_{{\rm rem},i}
    =
    \frac{3}{2}T_i+\frac{n}{4}T_i.
\end{equation}
For the saturated parametrization
$\sigma_i v_{\rm rel}\propto
[1+(v_{\rm rel}/v_0)^4]^{-1}$, the corresponding weighted average satisfies
\begin{equation}
    \frac{3}{4}T_i
    \leq
    \epsilon_{{\rm rem},i}
    \leq
    \frac{3}{2}T_i.
\end{equation}
Thus, in both parametrizations, the velocity-selection correction remains
of order $T_i$ and is suppressed relative to the leading inelastic energy
exchange $\Delta m$ by $T_i/\Delta m$. Since
$T_i/\Delta m\ll1$ during the relevant conversion epoch, we use
$\epsilon_{{\rm rem},i}\simeq3T_i/2$ in the leading-order temperature
evolution equations below.

Assuming the dark sector is closed, i.e., that there is no energy loss to
other sectors, energy conservation implies
$\mathrm{d}Q_h/\mathrm{d}t=-\mathrm{d}Q_l/\mathrm{d}t$.

Using an independent internal energy of each species $\mathrm{d}U_l = \mathrm{d}[ n_l V (m_l + 3T_l/2) ]$, we obtain 
\begin{equation}
    \frac{\mathrm{d}U_l}{\mathrm{d}t} = 
    V \left( m_l + \frac{3}{2}T_l \right) \dot{n}_l
    + \frac{3}{2} V n_l \dot{T}_l
    + \left( m_l n_l + \frac{3}{2} n_l T_l \right) \frac{\dot{V}}{V} V \,.
    \label{eq:Ul_dot}
\end{equation}
Using $\dot{V}/V = 3H$, substituting Eqs.~\eqref{eq:firstlaw}, \eqref{eq:Ql_dot}, and \eqref{eq:Ul_dot}, and utilizing the number density Eq.~\eqref{eq:nl_dot}, we arrive at the temperature evolution equation for $\chi_l$,
\begin{equation}
    \dot{T}_l = -2H T_l + \frac{2}{3} \langle\sigma v\rangle_{hh\to ll}(T_h) \frac{n_h^2}{n_l}
    \left[ \Delta m + \frac{3}{2}(T_h - T_l) \right] \, .
    \label{eq:Tl_dot}
\end{equation}
Similarly, for the heavy component $\chi_h$,
\begin{equation}
    \dot{T}_h = -2H T_h - \frac{2}{3} \langle\sigma v\rangle_{ll\to hh}(T_l) \frac{n_l^2}{n_h}
    \left[ \Delta m + \frac{3}{2}(T_h - T_l) \right] \,.
    \label{eq:Th_dot}
\end{equation}
Expressing these in terms of redshift $z$:
\begin{align}
    \frac{\mathrm{d}T_l}{\mathrm{d}z} &= \frac{2T_l}{1+z} 
    - \frac{2}{3(1+z)H(z)} \frac{n_h^2}{n_l} \langle\sigma v\rangle_{hh\to ll}(T_h)
    \left[ \Delta m + \frac{3}{2}(T_h - T_l) \right] \,,
    \label{eq:Tl_z} \\[8pt]
    \frac{\mathrm{d}T_h}{\mathrm{d}z} &= \frac{2T_h}{1+z} 
    + \frac{2}{3(1+z)H(z)} \frac{n_l^2}{n_h} \langle\sigma v\rangle_{ll\to hh}(T_l)
    \left[ \Delta m + \frac{3}{2}(T_h - T_l) \right] \,.
    \label{eq:Th_z}
\end{align}
These equations rely on the assumption of rapid self-thermalization, ensuring each species is described by a single temperature. The term proportional to $\Delta m$ in Eq.~\eqref{eq:Tl_dot} represents \textit{chemical heating}: the exothermic de-excitation $hh\to ll$ converts the mass splitting energy into kinetic energy of the light species. Conversely, the endothermic excitation $ll\to hh$ acts as a cooling term for the heavy component distribution.

\subsection{Derivation of the Linear Perturbation Equations}
\label{app:perturb_derivation}
Throughout this derivation, both DM species are treated as non-relativistic particles in the classical Maxwell--Boltzmann regime. 
The derivation follows the energy--momentum transfer formalism for interacting fluids, which has been widely used in studies of decaying, annihilating, and converting dark-sector components
\cite{Erickcek:2011us,Fan:2014zua,Erickcek:2015jza,Choi:2015bya,Redmond:2018xty,Kim:2024bdm}.
For each species $i\in\{l,h\}$, the phase-space distribution
$f_i(x^\mu,p^\mu)$ evolves according to the covariant Boltzmann equation,
\begin{equation}
    p^\mu \frac{\partial f_i}{\partial x^\mu}
    -\Gamma^\lambda_{\alpha\beta} p^\alpha p^\beta
    \frac{\partial f_i}{\partial p^\lambda}
    = C_i[f],
    \label{eq:bolzman_eqs}
\end{equation}
where $C_i[f]$ is the collision operator describing the inelastic conversion process $\chi_h\chi_h\leftrightarrow\chi_l\chi_l$.

The fluid variables are obtained from momentum integrals of the distribution function. 
In particular, the energy--momentum tensor of each species is given by
\begin{equation}
    {}^{(i)}T^{\mu\nu}
    \equiv
    g_i \int d\Pi_i  p^\mu p^\nu f_i,
    \qquad
    d\Pi_i \equiv \frac{d^3p}{(2\pi)^3 E_i},
\end{equation}
where $g_i$ is the internal degeneracy. 
In the ideal-fluid approximation it takes the  form
\begin{equation}
    {}^{(i)}T^{\mu\nu} =
    (\rho_i+p_i)\,u_i^\mu u_i^\nu + p_i\,g^{\mu\nu},
\end{equation}
with $\rho_i$, $p_i$, and $u_i^\mu$ the energy density, pressure, and four-velocity of the $i$-th component.

Taking the first momentum moment of the Boltzmann equation Eq.~\eqref{eq:bolzman_eqs} gives
\begin{equation}
    \nabla_\mu {}^{(i)}T^\mu{}\nu
    =
    g_i\int d\Pi_i,p_\nu C_i[f]
    \equiv Q_\nu^{(i)}, 
    \label{eq:app_conservation}
\end{equation}
with the source term $Q_\nu^{(i)}$ describes the energy--momentum transferred to species $i$ by the conversion process.

We now evaluate $Q_\nu^{(i)}$ for $\chi_h(p_1)\chi_h(p_2)\leftrightarrow\chi_l(p_3)\chi_l(p_4)$, taking the
light component as an example. For the light component,
\begin{align}
    Q_\nu^{(l)}
    &=
    \frac{1}{2}
    \int d\Pi_1 d\Pi_2 d\Pi_3 d\Pi_4\,
    (2\pi)^4
    \delta^{(4)}(p_1+p_2-p_3-p_4)
    |\mathcal M|^2
    (p_{3\nu}+p_{4\nu})
    \nonumber\\
    &\qquad\times
    \left[
        f_h(p_1)f_h(p_2)
        -
        f_l(p_3)f_l(p_4)
    \right],
    \label{eq:app_Ql_collision_full}
\end{align}
where the factor $1/2$ accounts for the identical two-particle state.

Using momentum conservation and the definition of $\sigma v_{\rm rel}$,
Eq.~\eqref{eq:app_Ql_collision_full} becomes
\begin{align}
    Q_\nu^{(l)}
    &=
    \frac{1}{2}
    \int
    \frac{g_h d^3p_1}{(2\pi)^3}
    \frac{g_h d^3p_2}{(2\pi)^3}
    f_h(p_1)f_h(p_2)\,
    \sigma_{hh\to ll}v_{\rm rel}\,
    (p_{1\nu}+p_{2\nu})
    \nonumber\\
    &\quad
    -
    \frac{1}{2}
    \int
    \frac{g_l d^3p_3}{(2\pi)^3}
    \frac{g_l d^3p_4}{(2\pi)^3}
    f_l(p_3)f_l(p_4)\,
    \sigma_{ll\to hh}v_{\rm rel}\,
    (p_{3\nu}+p_{4\nu}) .
    \label{eq:app_Ql_rate_weighted}
\end{align}

We now factorize the first term in Eq.~\eqref{eq:app_Ql_rate_weighted}. For the
incoming heavy pair, define the center-of-mass and relative velocities as
\begin{equation}
    \mathbf V=\frac{\mathbf v_1+\mathbf v_2}{2},
    \qquad
    \mathbf v_{\rm rel}=\mathbf v_1-\mathbf v_2 ,
\end{equation}
with $d^3v_1d^3v_2=d^3Vd^3v_{\rm rel}$. For a locally thermalized heavy
component with bulk velocity $\mathbf u_h$, the product of Maxwell--Boltzmann
distributions factorizes as
\begin{equation}
    f_h(\mathbf v_1)f_h(\mathbf v_2)
    \propto
    \exp\left[-\frac{m_h(\mathbf V-\mathbf u_h)^2}{T_h}\right]
    \exp\left[-\frac{m_h v_{\rm rel}^2}{4T_h}\right].
    \label{eq:app_distribution_factorization}
\end{equation}

Applying this change of variables, we obtain
\begin{align}
    Q_\nu^{(l)}\big|_{hh\to ll}
    &=
    \frac{1}{2} n_h^2
    \frac{
    \int e^{-m_h(\mathbf V-\mathbf u_h)^2/T_h}(p_{1\nu}+p_{2\nu}) d^3V ~        e^{-m_hv_{\rm rel}^2/4T_h}
    \sigma_{hh\to ll}v_{\rm rel} d^3v_{\rm rel}
    }{
    \int d^3V\,d^3v_{\rm rel}
    e^{-m_h(\mathbf V-\mathbf u_h)^2/T_h}
    e^{-m_hv_{\rm rel}^2/4T_h}
    } 
    \nonumber
    \\
    &= \frac{1}{2} n_h^2 
    \left\langle p_{1\nu}+p_{2\nu}\right\rangle_{\mathbf V}
    \langle\sigma v\rangle_{hh\to ll}
    .
    \label{eq:app_Q_hhll_V_vrel}
\end{align}
Here the normalization follows from
$n_h=g_h\int d^3p/(2\pi)^3 f_h(p)$.

The remaining center-of-mass average is evaluated from the single-particle four-momentum,
\begin{equation}
    p_{a0} =  -m_h,
    \qquad
    p_{ai}=m_h v_{a,i},
    \qquad a=1,2,
\end{equation}
where we use the metric convention $(-,+,+,+)$. Since the center-of-mass
distribution is centered at the bulk velocity $\mathbf u_h$, one has
$
    \langle 1\rangle_{\mathbf V}=1,
    \langle V_i\rangle_{\mathbf V}=u_{h,i}.
$
Therefore,
$
    \langle p_{10}+p_{20}\rangle_{\mathbf V}
    =
    -2m_h$, 
    $\langle p_{1i}+p_{2i}\rangle_{\mathbf V}
    =
    2m_h u_{h,i}.
$
Equivalently,
\begin{equation}
    \left\langle p_{1\nu}+p_{2\nu}\right\rangle_{\mathbf V}
    =
    2m_h u_\nu^{(h)} .
\label{eq:app_pair_momentum_average}
\end{equation}
Substituting Eq.~\eqref{eq:app_pair_momentum_average} into
Eq.~\eqref{eq:app_Q_hhll_V_vrel} gives
\begin{equation}
    Q_\nu^{(l)}\big|_{hh\to ll}
    =
    m_h n_h^2
    \langle\sigma v\rangle_{hh\to ll}
    u_\nu^{(h)}
    =
    \langle\sigma v\rangle_{hh\to ll}
    \frac{\rho_h^2}{m_h}
    u_\nu^{(h)} ,
    \label{eq:app_Q_hhll_rho}
\end{equation}
where we used $\rho_h\simeq m_h n_h$ for non-relativistic particles.

The $Q_\nu^{(l)}\big|_{ll\to hh}$ is obtained analogously and carries the opposite
sign for the light component. Combining the two contributions, we obtain
\begin{equation}
Q_\nu^{(l)}
=
\langle\sigma v\rangle_{hh\to ll}
\frac{\rho_h^2}{m_h}
u_\nu^{(h)}
-
\langle\sigma v\rangle_{ll\to hh}
\frac{\rho_l^2}{m_l}
u_\nu^{(l)} .
\label{eq:app_Ql_form}
\end{equation}

Having obtained the covariant source term $Q_\nu^{(i)}$, we now derive the
linear perturbation equations by expanding the energy--momentum transfer
equations around the homogeneous background. We work in Newtonian gauge,
which in cosmic time $t$ reads
\begin{equation}
    ds^2 = -(1+2\Psi)\,dt^2 + a^2(t)\,(1-2\Phi)\,d\mathbf{x}^2,
    \label{eq:app_metric}
\end{equation}
where $\Psi$ and $\Phi$ are the metric potentials and $H\equiv\dot a/a$ is the
Hubble rate, with a dot denoting $d/dt$.

For each species, perturbations are defined by
\begin{equation}
    \rho_i(t,\mathbf{x})=\bar\rho_i(t)\,[1+\delta_i(t,\mathbf{x})],
    \qquad
    \theta_i \equiv \partial_j v_i^j,
    \label{eq:app_delta_theta_def}
\end{equation}
where $v_i^j$ is the peculiar velocity. We normalize the four-velocity in the
cosmic-time frame, so that the background timelike component is
$u_0^{(0)}=-1$, consistent with evaluating the transfer components $Q_\nu$ at
$|_t$. To first order,
\begin{equation}
    u_i^0=1-\Psi,\qquad
    u_i^j=\frac{1}{a}v_i^j,
    \qquad
    u_{i0}=-(1+\Psi),\qquad
    u^{(i)}_j\simeq a v^{(i)}_j.
    \label{eq:app_u_expansion}
\end{equation}

Following Refs.~\cite{Redmond:2018xty,Erickcek:2015jza}, the perturbed conservation equations~\eqref{eq:app_conservation} for non-relativistic species ($w_i=0$) yield the continuity and Euler equations
\begin{align}
    \frac{d\delta_i}{dt} + \frac{\theta_i}{a} - 3\frac{d\Phi}{dt}
    &= \frac{1}{\bar\rho_i}\Big(\bar Q^{(i)}_{0}\big|_t\,\delta_i
    - Q^{(i,1)}_{0}\big|_t\Big),
    \label{eq:app_delta_master}\\
    \frac{d\theta_i}{dt} + H\theta_i - \frac{k^2\Psi}{a}
    - \frac{\delta P_i}{\delta\rho_i}\frac{k^2\delta_i}{a}
    &= \frac{1}{\bar\rho_i}\bigg(
    \frac{\partial_j Q^{(i,1)}_{j}\big|_t}{a}
    + \bar Q^{(i)}_{0}\big|_t\,\theta_i\bigg),
    \label{eq:app_theta_master}
\end{align}
where $|_t$ indicates that the transfer components are evaluated with $u_0^{(0)}=-1$.

Expanding Eq.~\eqref{eq:app_Ql_form} to first order using
$\rho_i = \bar\rho_i(1+\delta_i)$, the relevant components are
\begin{align}
    \bar Q^{(l)}_{0}\big|_t
    &= -\avg{\sigma v}_{hh\to ll}\frac{\bar\rho_h^2}{m_h}
    + \avg{\sigma v}_{ll\to hh}\frac{\bar\rho_l^2}{m_l},
    \label{eq:app_Q0_bg}\\
    Q^{(l,1)}_{0}\big|_t
    &= -\avg{\sigma v}_{hh\to ll}\frac{\bar\rho_h^2}{m_h}(2\delta_h+\Psi)
    + \avg{\sigma v}_{ll\to hh}\frac{\bar\rho_l^2}{m_l}(2\delta_l+\Psi),
    \label{eq:app_Q0_1}\\
    \frac{\partial_j Q^{(l,1)}_{j}\big|_t}{a}
    &= \avg{\sigma v}_{hh\to ll}\frac{\bar\rho_h^2}{m_h}\,\theta_h
    - \avg{\sigma v}_{ll\to hh}\frac{\bar\rho_l^2}{m_l}\,\theta_l.
    \label{eq:app_Qi_1}
\end{align}

Substituting Eqs.~\eqref{eq:app_Q0_bg}--\eqref{eq:app_Qi_1} into
Eqs.~\eqref{eq:app_delta_master}--\eqref{eq:app_theta_master} and
converting to conformal time using $d/dt=(1/a) d/d\tau$,  $H=\mathcal{H}/a$, and the energy density ratio $\xi \equiv \bar\rho_h/\bar\rho_l$, which is related to the number fraction $r_h \equiv n_h/(n_l+n_h)$ by $\xi = (m_h/m_l) r_h/(1-r_h)$.
The resulting conformal-time perturbation equations are
\begin{align}
    \delta_l' &=
    -\theta_l + 3\Phi'
    + \xi\mathcal{R}_{d}\Big(2\delta_h-\delta_l+\Psi\Big)
    -\mathcal{R}_{u}\Big(\delta_l+\Psi\Big),
    \label{eq:app_delta_l}\\
    \delta_h' &=
    -\theta_h + 3\Phi'
    + \frac{1}{\xi}\mathcal{R}_{u}\Big(2\delta_l-\delta_h+\Psi\Big)
    -\mathcal{R}_{d}\Big(\delta_h+\Psi\Big),
    \label{eq:app_delta_h}
\end{align}
and
\begin{align}
    \theta_l' &=
    -\mathcal{H}\theta_l + k^2\Psi + k^2 c_{s,l}^2\,\delta_l
    + \xi\mathcal{R}_d\,(\theta_h-\theta_l),
    \label{eq:app_theta_l}\\
    \theta_h' &=
    -\mathcal{H}\theta_h + k^2\Psi
    + \frac{1}{\xi}\mathcal{R}_u\,(\theta_l-\theta_h).
    \label{eq:app_theta_h}
\end{align}
Equations~\eqref{eq:app_delta_l}--\eqref{eq:app_theta_h} reduce to the
standard CDM limit when $\mathcal{R}_d=\mathcal{R}_u=0$.

\subsection{Second-Order Density Perturbation Equations}
\label{app:second_order}

To analyse the physical regimes discussed in the main text it is useful to
have closed second-order equations for $\delta_l$ and $\delta_h$ in which the
velocity divergences have been eliminated. We set $\mathcal{R}_u = 0$
throughout.
\begin{itemize}
\item \textbf{Heavy component}

From Eq.~\eqref{eq:app_delta_h} we extract
\begin{equation}
    \theta_h = -\delta_h' + 3\Phi' - \mathcal{R}_d(\delta_h + \Psi).
    \label{eq:app_theta_h_expr}
\end{equation}
Differentiating Eq.~\eqref{eq:app_delta_h} and substituting
Eq.~\eqref{eq:app_theta_h} for $\theta_h'$, then using
Eq.~\eqref{eq:app_theta_h_expr} to eliminate $\theta_h$, we obtain
\begin{equation}
    \delta_h'' + (\mathcal{H} + \mathcal{R}_d)\,\delta_h'
    + (\mathcal{R}_d\mathcal{H} + \mathcal{R}_d')\,\delta_h
    = -k^2\Psi + S_h,
    \label{eq:app_delta_h_2nd}
\end{equation}
where the source term depending only on the metric potentials is
\begin{equation}
    S_h = -(\mathcal{R}_d\mathcal{H} + \mathcal{R}_d')\,\Psi
    - \mathcal{R}_d\,\Psi'
    + 3\Phi'' + 3\mathcal{H}\Phi'.
    \label{eq:app_Sh}
\end{equation}
The friction coefficient $\mathcal{H}+\mathcal{R}_d$ exceeds the standard
Hubble drag by the conversion rate, and the term
$(\mathcal{R}_d\mathcal{H}+\mathcal{R}_d')\,\delta_h$ acts as an effective
dissipation of the heavy-component overdensity.

\item \textbf{Light component}

From Eq.~\eqref{eq:app_delta_l} we extract
\begin{equation}
    \theta_l = -\delta_l' + 3\Phi'
    + \xi\mathcal{R}_d(2\delta_h - \delta_l + \Psi).
    \label{eq:app_theta_l_expr}
\end{equation}
The velocity difference follows from
Eqs.~\eqref{eq:app_theta_h_expr} and \eqref{eq:app_theta_l_expr},
\begin{equation}
    \theta_h - \theta_l = \delta_l' - \delta_h'
    - \mathcal{R}_d(\delta_h+\Psi)
    - \xi\mathcal{R}_d(2\delta_h - \delta_l + \Psi).
    \label{eq:app_theta_diff}
\end{equation}
Differentiating Eq.~\eqref{eq:app_delta_l} and substituting
Eq.~\eqref{eq:app_theta_l} for $\theta_l'$, then using
Eqs.~\eqref{eq:app_theta_l_expr} and \eqref{eq:app_theta_diff} to
eliminate $\theta_l$ and $\theta_h-\theta_l$, we obtain after collecting terms
\begin{equation}
    \delta_l'' + (\mathcal{H} + 2\xi\mathcal{R}_d)\,\delta_l'
    + \Big[\xi\mathcal{R}_d\mathcal{H} + (\xi\mathcal{R}_d)^2
    + (\xi\mathcal{R}_d)' + k^2 c_s^2\Big]\,\delta_l
    = -k^2\Psi + S_l,
    \label{eq:app_delta_l_2nd}
\end{equation}
where
\begin{align}
    S_l &= \big(\xi\mathcal{R}_d\mathcal{H} + \xi\mathcal{R}_d^2
    + (\xi\mathcal{R}_d)^2 + (\xi\mathcal{R}_d)'\big)\,\Psi
    + \xi\mathcal{R}_d\,\Psi'
    + 3\mathcal{H}\Phi' + 3\Phi''
    \nonumber\\
    &\quad + 3\xi\mathcal{R}_d\,\delta_h'
    + \big(\xi\mathcal{R}_d^2 + 2(\xi\mathcal{R}_d)^2
    + 2(\xi\mathcal{R}_d)' + 2\xi\mathcal{R}_d\mathcal{H}\big)\,\delta_h.
    \label{eq:app_Sl}
\end{align}

Several features of Eq.~\eqref{eq:app_delta_l_2nd} are worth noting.
The friction coefficient $\mathcal{H}+2\xi\mathcal{R}_d$ contains a conversion
contribution $2\xi\mathcal{R}_d$ that is twice that appearing in the heavy
equation. The additional factor arises from eliminating the velocity
difference $\theta_h-\theta_l$.
The effective frequency squared of the $\delta_l$ mode receives contributions
from both the sound speed ($k^2 c_s^2$) and the conversion process
($\xi\mathcal{R}_d\mathcal{H}+(\xi\mathcal{R}_d)^2+(\xi\mathcal{R}_d)'$).
The latter are independent of $k$ and arise purely from the inelastic
scattering.
The source $S_l$ couples $\delta_l$ to $\delta_h$ and $\delta_h'$,
representing the injection of converted particles from the heavy into the
light component.

\end{itemize}

\subsection{Sound Speed and Pressure Perturbations in the Light Sector}
\label{app:sound_speed}

In the derivation of the momentum conservation equation for the light DM state $\chi_l$, we introduced a pressure gradient term proportional to an effective sound speed squared, $c_{s,l}^2$. Given that the exothermic conversion $hh \to ll$ significantly alters the thermal history of the light sector, causing it to deviate from standard adiabatic cooling, care must be taken in defining the sound speed. The adiabatic sound speed, $c_{a,l}^2$, is strictly defined by the evolution of the background pressure $\bar{P}_l$ and density $\bar{\rho}_l$ as $c_{a,l}^2 \equiv \dot{\bar{P}}_l / \dot{\bar{\rho}}_l$. Assuming the light component maintains local thermal equilibrium and behaves as a non-relativistic ideal gas , this definition yields 
\begin{equation}
    c_{a,l}^2(\tau) = \frac{T_l}{m_l} \left( 1 - \frac{1}{3} \frac{d \ln T_l}{d \ln a} \right) \,.
    \label{eq:ca2_formula}
\end{equation}
This expression governs the evolution of the fluid stiffness, 
incorporating chemical heating effects through the modified temperature slope $d \ln T_l / d \ln a$.

The sound speed appearing in the perturbation equations, 
often called as the effective sound speed $c_{\rm eff}^2 \equiv \delta P_l / \delta \rho_l$, 
is generally frame-dependent. 
It is physically useful to relate $c_{\rm eff}^2$ to the rest-frame sound speed $\hat{c}_{s,l}^2$, 
which describes the intrinsic pressure response to density fluctuations. 
In the Newtonian gauge, this relation is given by a gauge transformation of the pressure perturbation $\delta P_l$, 
\begin{equation}
    \delta P = c_a^2 \delta \rho + (\hat{c_s}^2 - c_a^2) (\delta \rho - \dot{\rho} \frac{\theta}{k^2}).
    \label{eq:deltaP_gauge}
\end{equation}
This implies that the effective sound speed in the calculation frame differs from the rest-frame value by a term proportional to $(\hat{c}_{s,l}^2 - c_{a,l}^2)$ and the velocity divergence.

In our analysis, we adopt the standard fluid approximation by identifying the effective sound speed with the adiabatic sound speed, $c_{s,l}^2 \approx \hat{c}_{s,l}^2 \approx c_{a,l}^2$. 
This approximation assumes local thermal equilibrium: despite vigorous energy injection from the heavy sector, 
the light DM particles maintain (or rapidly relax to) a local Maxwell-Boltzmann distribution with temperature 
$T(\mathbf{x}) = \bar{T} + \delta T$. 
Under this condition, the rest-frame pressure perturbation follows the background instantaneous adiabatic relation.

Any deviations from this adiabatic response (i.e., intrinsic entropy perturbations) are treated as sub-dominant in the fluid limit. Furthermore, distinct from the intrinsic sound speed of the light component, the \textit{relative entropy perturbations} between the heavy and light sectors (arising from their distinct evolution) are explicitly governed by the interaction terms in the coupled Boltzmann equations (Eqs.~\eqref{eq:delta_l}--\eqref{eq:theta_l}). 
Consequently, the pressure gradient term in Eq.~\eqref{eq:theta_l} describes 
the internal hydrodynamical dynamics of the light fluid, 
consistently measured by the adiabatic sound speed in Eq.~\eqref{eq:ca2_formula}.

\bibliographystyle{JHEP}
\bibliography{reference.bib}

\end{document}